\documentclass[11pt]{article}
\usepackage{amsmath}
\usepackage{ amssymb, amsthm, amsfonts, mathrsfs, amsbsy,  wasysym }
\usepackage[T1]{fontenc}
\theoremstyle{definition}
\usepackage{fullpage}
\usepackage[utf8]{inputenc}

\usepackage{enumerate}
\usepackage{graphicx}
\usepackage{mathtools} 
\usepackage{float}
\usepackage{extpfeil}

\usepackage{amssymb}
\usepackage{amsmath}
\usepackage{physics}
\usepackage{nicefrac}

\usepackage{subcaption}
\usepackage[export]{adjustbox}
\usepackage{stackrel}
\usepackage[usenames,dvipsnames]{xcolor}

\usepackage[skins]{tcolorbox}
\tcbuselibrary{breakable}
\usepackage{ textcomp }

\usepackage[all,cmtip]{xy}

\usepackage{circledsteps}
\usepackage{hyperref}

\usepackage{listings}

\usepackage{stackengine}
\usepackage{bbm}
\usepackage{authblk}

\usetikzlibrary{arrows.meta, positioning}
\usepackage[
  backend=biber,
  style=authoryear,
  maxcitenames=2,
  maxbibnames=99,
  url=false,
]{biblatex}
\usepackage{pdfpages}
\usepackage{multirow}
\usepackage{makecell}

\stackMath
\newcommand\tsup[2][2]{%
 \def\useanchorwidth{T}%
  \ifnum#1>1%
    \stackon[-.5pt]{\tsup[\numexpr#1-1\relax]{#2}}{\scriptscriptstyle\sim}%
  \else%
    \stackon[.5pt]{#2}{\scriptscriptstyle\sim}%
  \fi%
}

\makeatletter
\newcommand{\leqnomode}{\tagsleft@true}
\newcommand{\reqnomode}{\tagsleft@false}
\makeatother

\SetSymbolFont{stmry}{bold}{U}{stmry}{m}{n}

\newtcolorbox{mybox}[2]{enhanced,sharp corners=all,colback=white,colframe=white,toprule=0.0cm,bottomrule=0pt,leftrule=0pt,rightrule=0pt,grow to left by=15pt, left=14pt, grow to right by=15pt, right=14pt, overlay={
            \draw[darkgray,line width=2pt] (frame.north west)+(-.5pt,0pt) -- ++(2cm,0pt);
            \draw[darkgray,line width=2pt] (frame.south east)+(1pt,0pt) -- ++(-2cm,0pt);
            \draw[darkgray,line width=2pt] (frame.south west) -- (frame.north west)+(0pt,1pt);
            \draw[darkgray,line width=2pt] (frame.north east) -- (frame.south east)+(0pt,-1pt);
    },
    coltitle=black,colbacktitle=white,titlerule=0pt,
    title={\vskip5pt\bfseries{\underline{#1}}\hfill \bfseries{#2}}
}

\newtcolorbox{mylemmabox}[2]{enhanced,sharp corners=all,colback=white,colframe=white,toprule=0.0cm,bottomrule=0pt,leftrule=0pt,rightrule=0pt,grow to left by=15pt, left=14pt, grow to right by=15pt, right=14pt, overlay={
            \draw[darkgray,line width=2pt] (frame.north west)+(-.5pt,0pt) -- ++(2cm,0pt);
            \draw[darkgray,line width=2pt] (frame.south east)+(1pt,0pt) -- ++(-2cm,0pt);
            \draw[darkgray,line width=2pt] (frame.south west) -- (frame.north west)+(0pt,1pt);
            \draw[darkgray,line width=2pt] (frame.north east) -- (frame.south east)+(0pt,-1pt);
    },
    coltitle=black,colbacktitle=white,titlerule=0pt,
    title={\vskip5pt\bfseries{\underline{Lemma #1:}}\hfill \bfseries{#2}}
}

\newtcolorbox{mybox0}[2]{enhanced,sharp corners=all,colback=white,colframe=white,toprule=0.0cm,bottomrule=0pt,leftrule=0pt,rightrule=0pt,grow to left by=15pt, left=14pt, grow to right by=15pt, right=14pt, overlay={
            \draw[gray,line width=2pt] (frame.north west) -- ++(2cm,0pt);
            \draw[gray,line width=2pt] (frame.south east) -- ++(-2cm,0pt);
            \draw[gray,line width=2pt] (frame.south west) -- (frame.north west);
            \draw[gray,line width=2pt] (frame.north east) -- (frame.south east);
    },
    coltitle=black,colbacktitle=white,titlerule=0pt,
    title={\vskip5pt\bfseries{\underline{Theorem #1:}}\hfill \bfseries{(#2)}}
}

\newtcolorbox{mybox1}{enhanced,sharp corners=all,colback=white,colframe=white,toprule=0cm,bottomrule=0pt,leftrule=0pt,rightrule=0pt,grow to left by=15pt, left=14pt, grow to right by=15pt, right=14pt, overlay={
            \draw[gray,line width=2pt] (frame.north west)+(0cm,2.5mm) -- (frame.south west);
    },
}

\newtcolorbox{mybox2}{enhanced,sharp corners=all,colback=white,colframe=white,toprule=0cm,bottomrule=0pt,leftrule=0pt,rightrule=0pt,grow to left by=15pt, left=14pt, grow to right by=15pt, right=14pt, overlay={
            \draw[gray,line width=2pt] (frame.north west)+(0cm,2.5mm) -- (frame.south west);
    },
}

\newtcolorbox{mybox3}{enhanced,sharp corners=all,colback=white,colframe=white,toprule=0cm,bottomrule=0pt,leftrule=0pt,rightrule=0pt,grow to left by=15pt, left=14pt, grow to right by=15pt, right=14pt, overlay={
            \draw[darkgray,line width=2pt] (frame.north west)+(0cm,2.5mm) -- (frame.south west);
            \draw[darkgray,line width=2pt] (frame.north west)+(0cm,-13pt) -- ++(5mm,-13pt);
            \draw[darkgray,line width=2pt] (frame.south west) -- (frame.south east)+(-10pt,0pt);
    },
}

\newtcolorbox{mybox4}[2]{enhanced,sharp corners=all,colback=white,colframe=gray,toprule=2pt,bottomrule=2pt,leftrule=0pt,rightrule=2pt,grow to left by=15pt, left=14pt, grow to right by=15pt, right=14pt, overlay={            \draw[gray,line width=2pt] (frame.south west) -- (frame.north west);},
    coltitle=black,colbacktitle=pink!15!,titlerule=1pt,
    title={\vskip5pt\bfseries{#1}\hfill \bfseries{#2}}
}

\newtcolorbox{ultracustom}[4]{enhanced,sharp corners=all,colback=white,colframe=#3,toprule=2pt,bottomrule=2pt,leftrule=0pt,rightrule=2pt,grow to left by=15pt, left=14pt, grow to right by=15pt, right=14pt, overlay={            \draw[#3,line width=2pt] (frame.south west) -- (frame.north west);},
    coltitle=black,colbacktitle=#4!15!,titlerule=1pt,
    title={\vskip5pt\bfseries{#1}\hfill \bfseries{#2}}
}

\newenvironment{customnumbered}[1][]{%
  \refstepcounter{theo}
  \begin{ultracustom}{Proposition \thetheo.}{#1}{darkgray}{gray}%
}{%
  \end{ultracustom}%
}
\newenvironment{customnumbered2}[1][]{%
  \refstepcounter{theo}
  \begin{ultracustom}{Theorem \thetheo.}{#1}{darkgray}{gray}%
}{%
  \end{ultracustom}%
}

\newtcolorbox{mybox5}{enhanced,sharp corners=all,colback=white,colframe=white,toprule=0cm,bottomrule=0pt,leftrule=0pt,rightrule=0pt,grow to left by=15pt, left=14pt, grow to right by=15pt, right=14pt, overlay={
            \draw[gray,line width=2pt] (frame.north west)+(0cm,-12.5pt) -- (frame.south west);
            \draw[gray,line width=2pt] (frame.south west) -- (frame.south east)+(-10pt,0pt);
            \draw[gray,line width=2pt] (frame.north west)+(0cm,-13pt) -- ++(5mm,-13pt);
    },
}

\newtcolorbox{customproof}[1]{enhanced,breakable,sharp corners=all,colback=white,colframe=white,toprule=0cm,bottomrule=0pt,leftrule=0pt,rightrule=0pt,grow to left by=15pt, left=14pt, grow to right by=15pt, right=14pt,
    overlay unbroken={
		\draw[#1,line width=2pt] (frame.north west)+(0cm,-12.5pt) -- (frame.south west);
    \draw[#1,line width=2pt] (frame.south west) -- (frame.south east);
    \draw[#1,line width=2pt] (frame.north west)+(0cm,-13pt) -- ++(5mm,-13pt);
	},
	overlay first={
		\draw[#1,line width=2pt] (frame.north west)+(0cm,-13pt) -- ++(5mm,-13pt);
		\draw[#1,line width=2pt] (frame.north west)+(0cm,-12.5pt) -- (frame.south west);
	},
	overlay middle={
		\draw[#1,line width=2pt] (frame.north west) -- (frame.south west);
	},
	overlay last={
		\draw[#1,line width=2pt] (frame.north west) -- (frame.south west);
		\draw[#1,line width=2pt] (frame.south west) -- (frame.south east);
	}
}

\newtcolorbox{mybox55}{enhanced,breakable,sharp corners=all,colback=white,colframe=white,toprule=0cm,bottomrule=0pt,leftrule=0pt,rightrule=0pt,grow to left by=15pt, left=14pt, grow to right by=15pt, right=14pt, 
					extras unbroken and last={borderline south={2pt}{0pt}{gray}},
					overlay={
            \draw[gray,line width=2pt] (frame.north west)+(0cm,-12.5pt) -- (frame.south west);
            \draw[gray,line width=2pt] (frame.south west) -- (frame.south east)+(-10pt,0pt);
            \draw[gray,line width=2pt] (frame.north west)+(0cm,-13pt) -- ++(5mm,-13pt);
    },
}

\newtcolorbox{mybox6}{enhanced,sharp corners=all,colback=white,colframe=white,toprule=0cm,bottomrule=0pt,leftrule=0pt,rightrule=0pt,grow to left by=15pt, left=14pt, grow to right by=15pt, right=14pt, overlay={
            \draw[gray,line width=2pt] (frame.north west)+(0cm,-12.5pt) -- (frame.south west)+(0cm,12pt);
            \draw[gray,line width=2pt] (frame.south west)+(0,15pt) -- ++(5mm,15pt);
            \draw[gray,line width=2pt] (frame.north west)+(0cm,-13pt) -- ++(5mm,-13pt);
            \draw[white,line width=3pt] (frame.south west) -- ++(0cm,14pt);
    },
}

\newtcolorbox{mybox7}{enhanced,sharp corners=all,colback=white,colframe=white,toprule=0cm,bottomrule=0pt,leftrule=0pt,rightrule=0pt,grow to left by=15pt, left=14pt, grow to right by=15pt, right=14pt, overlay={
            \draw[gray,line width=2pt] (frame.north west)+(0cm,-12.5pt) -- (frame.south west)+(0cm,12pt);
            \draw[gray,line width=2pt] (frame.north west)+(0cm,-13pt) -- ++(5mm,-13pt);
    },
}

\newtcolorbox{mybox8}[2]{enhanced,breakable,sharp corners=all,colback=white,colframe=white,toprule=0.0cm,bottomrule=0pt,leftrule=0pt,rightrule=0pt,grow to left by=15pt, left=14pt, grow to right by=15pt, right=14pt, 
					extras unbroken and last={
						borderline south={2pt}{0pt}{gray}
						},
					overlay={
						\ifcase\tcbsegmentstate
							\draw[gray,line width=2pt] (frame.south west) -- (frame.north west)+(0pt,1pt);
							\draw[gray,line width=2pt] (frame.north east) -- (frame.south east)+(0pt,-1pt);
							\draw[gray,line width=2pt] (frame.north west)+(-.5pt,0pt) -- ++(2cm,0pt);
							\or
							\draw[gray,line width=2pt] (frame.south west) -- (frame.north west)+(0pt,1pt);
							\draw[gray,line width=2pt] (frame.north east) -- (segmentation.east)+(0pt,-1pt);
							\draw[gray,line width=2pt] (frame.north west)+(-.5pt,0pt) -- ++(2cm,0pt);
							\draw[gray,line width=2pt] (segmentation.east)+(1pt,0pt) -- ++(-2cm,0pt);
							\draw[gray,line width=2pt] (segmentation.west)+(-.5pt,-13pt) -- ++(5mm,-13pt);
							\else
							\draw[gray,line width=2pt] (frame.south west) -- (frame.north west)+(0pt,1pt);
						\fi
    },
    coltitle=black,colbacktitle=white,titlerule=0pt,
    title={\vskip5pt\bfseries{\underline{Theorem #1:}}\hfill \bfseries{(#2)}}
}

\newtcolorbox{zProof}{
	enhanced,
	breakable,
	frame hidden,
	colback=white,
	grow to left by=15pt,
	left=14pt,
	grow to right by=15pt, 
	right=14pt,
	overlay unbroken={
		\draw[darkgray,line width=2pt] (frame.north west)+(0cm,-12.5pt) -- (frame.south west);
    \draw[darkgray,line width=2pt] (frame.south west) -- (frame.south east);
    \draw[darkgray,line width=2pt] (frame.north west)+(0cm,-13pt) -- ++(5mm,-13pt);
	},
	overlay first={
		\draw[darkgray,line width=2pt] (frame.north west)+(0cm,-13pt) -- ++(5mm,-13pt);
		\draw[darkgray,line width=2pt] (frame.north west)+(0cm,-12.5pt) -- (frame.south west);
	},
	overlay middle={
		\draw[darkgray,line width=2pt] (frame.north west) -- (frame.south west);
	},
	overlay last={
		\draw[darkgray,line width=2pt] (frame.north west) -- (frame.south west);
		\draw[darkgray,line width=2pt] (frame.south west) -- (frame.south east);
	}
}

\newtcolorbox{zContradict}{
	enhanced,
	enforce breakable,
	frame hidden,
	colback=white,
	grow to right by=15pt, 
	right=14pt,
    overlay unbroken={
		\draw[darkgray,line width=2pt] (frame.north west) -- (frame.south west);
	},
	overlay first={
		\draw[darkgray,line width=2pt] (frame.north west) -- (frame.south west);
	},
	overlay middle={
		\draw[darkgray,line width=2pt] (frame.north west) -- (frame.south west);
	},
	overlay last={
		\draw[darkgray,line width=2pt] (frame.north west) -- (frame.south west);
	}
}

\newtcolorbox{zzContradict}{
	enhanced,
	breakable,
	frame hidden,
	colback=white,
	grow to left by=15pt,
	grow to right by=15pt, 
	right=14pt,
	overlay ={
		\draw[darkgray,line width=2pt] (frame.north west) -- (frame.south west);
	}
}

\newtcolorbox{zzzTheo}{
	enhanced,
	breakable,
	frame hidden,
	segmentation hidden,
	segmentation at break=false,
	colback=white,
	grow to left by=15pt,
	left=14pt,
	grow to right by=15pt, 
	right=14pt,
	overlay last={
						\ifcase\tcbsegmentstate
							\draw[gray,line width=2pt] (frame.south west) -- (frame.north west)+(0pt,1pt);
							\draw[gray,line width=2pt] (frame.north east) -- (frame.south east)+(0pt,-1pt);
							\draw[gray,line width=2pt] (frame.north west)+(-.5pt,0pt) -- ++(2cm,0pt);
							\draw[gray,line width=2pt] (frame.south west) -- (frame.south east);
							\or
							\draw[gray,line width=2pt] (frame.south west) -- (frame.north west)+(0pt,1pt);
							\draw[gray,line width=2pt] (frame.north east) -- (segmentation.east)+(0pt,-1pt);
							\draw[gray,line width=2pt] (segmentation.east)+(1pt,0pt) -- ++(-2cm,0pt);
							\draw[gray,line width=2pt] (segmentation.west)+(-.5pt,-13pt) -- ++(5mm,-13pt);
							\draw[gray,line width=2pt] (frame.south west) -- (frame.south east);
							\else
							\draw[gray,line width=2pt] (frame.south west) -- (frame.north west)+(0pt,1pt);
							\draw[gray,line width=2pt] (frame.south west)+(-1pt,0pt) -- (frame.south east);
						\fi
						},
	overlay unbroken={
							\draw[gray,line width=2pt] (frame.south west) -- (frame.north west)+(0pt,1pt);
							\draw[gray,line width=2pt] (frame.north east) -- (segmentation.east)+(0pt,-1pt);
							\draw[gray,line width=2pt] (frame.north west)+(-.5pt,0pt) -- ++(2cm,0pt);
							\draw[gray,line width=2pt] (segmentation.east)+(1pt,0pt) -- ++(-2cm,0pt);
							\draw[gray,line width=2pt] (segmentation.west)+(-.5pt,-13pt) -- ++(5mm,-13pt);
							\draw[gray,line width=2pt] (frame.south west)+(-1pt,0pt) -- (frame.south east);
							},
	overlay first and middle={
						\ifcase\tcbsegmentstate
							\draw[gray,line width=2pt] (frame.south west) -- (frame.north west)+(0pt,1pt);
							\draw[gray,line width=2pt] (frame.north east) -- (frame.south east)+(0pt,-1pt);
							\draw[gray,line width=2pt] (frame.north west)+(-.5pt,0pt) -- ++(2cm,0pt);
							\or
							\draw[gray,line width=2pt] (frame.south west) -- (frame.north west)+(0pt,1pt);
							\draw[gray,line width=2pt] (frame.north east) -- (segmentation.east)+(0pt,-1pt);
							\draw[gray,line width=2pt] (frame.north west)+(-.5pt,0pt) -- ++(2cm,0pt);
							\draw[gray,line width=2pt] (segmentation.east)+(1pt,0pt) -- ++(-2cm,0pt);
							\draw[gray,line width=2pt] (segmentation.west)+(-.5pt,-13pt) -- ++(5mm,-13pt);
							\else
							\draw[gray,line width=2pt] (frame.south west) -- (frame.north west)+(0pt,1pt);
						\fi
						}		
}

\newtcolorbox{zTheoBox}{
	enhanced,
	breakable,
	frame hidden,
	colback=white,
	grow to left by=15pt,
	left=14pt,
	grow to right by=15pt, 
	right=14pt,
	overlay unbroken={
							\draw[gray,line width=2pt] (frame.south west) -- (frame.north west)+(0pt,1pt);
							\draw[gray,line width=2pt] (frame.north east) -- (frame.south east)+(0pt,-1pt);
							\draw[gray,line width=2pt] (frame.north west)+(-.5pt,0pt) -- ++(2cm,0pt);
							\draw[gray,line width=2pt] (frame.south east)+(1pt,0pt) -- ++(-2cm,0pt);
							},
	overlay first={
		\draw[gray,line width=2pt] (frame.north west)+(-.5pt,0pt) -- ++(2cm,0pt);
		\draw[gray,line width=2pt] (frame.north west) -- (frame.south west);
		\draw[gray,line width=2pt] (frame.north east) -- (frame.south east)+(0pt,-1pt);
	},
	overlay middle={
		\draw[gray,line width=2pt] (frame.north west) -- (frame.south west);
		\draw[gray,line width=2pt] (frame.north east) -- (frame.south east)+(0pt,-1pt);
	},
	overlay last={
		\draw[gray,line width=2pt] (frame.north west) -- (frame.south west);
		\draw[gray,line width=2pt] (frame.north east) -- (frame.south east)+(0pt,-1pt);
		\draw[gray,line width=2pt] (frame.south east)+(1pt,0pt) -- ++(-2cm,0pt);
	}
}

\makeatletter

\newcommand*{\relrelbarsep}{.386ex}
\newcommand*{\relrelbar}{%
	\mathrel{%
		\mathpalette\@relrelbar\relrelbarsep
	}%
}
\newcommand*{\@relrelbar}[2]{%
	\raise#2\hbox to 0pt{$\m@th#1\relbar$\hss}%
	\lower#2\hbox{$\m@th#1\relbar$}%
}
\providecommand*{\rightrightarrowsfill@}{%
	\arrowfill@\relrelbar\relrelbar\rightrightarrows
}
\providecommand*{\leftleftarrowsfill@}{%
	\arrowfill@\leftleftarrows\relrelbar\relrelbar
}
\providecommand*{\xrightrightarrows}[2][]{%
	\ext@arrow 0359\rightrightarrowsfill@{#1}{#2}%
}
\providecommand*{\xleftleftarrows}[2][]{%
	\ext@arrow 3095\leftleftarrowsfill@{#1}{#2}%
}

\usepackage{tikz}
\newif\ifCancelX
\tikzset{X/.code={\CancelXtrue}}
\newcommand{\Cancel}[2][]{\relax
\ifmmode%
  \tikz[baseline=(X.base),inner sep=0pt] {\node (X) {$#2$};
  \tikzset{#1}
  \draw[#1,overlay,shorten >=-2pt,shorten <=-2pt] (X.south west) -- (X.north east);
  \ifCancelX
    \draw[#1,overlay,shorten >=-2pt,shorten <=-2pt] (X.north west) -- (X.south east);   
  \fi}
\else
  \tikz[baseline=(X.base),inner sep=0pt] {\node (X) {#2};
  \tikzset{#1}
  \draw[#1,overlay,shorten >=-2pt,shorten <=-2pt] (X.south west) -- (X.north east);
  \ifCancelX
    \draw[#1,overlay,shorten >=-2pt,shorten <=-2pt] (X.north west) -- (X.south east);
  \fi}%
\fi}

\makeatletter
\DeclareRobustCommand{\butthole}{%
  \mathbin{\mathpalette\o@plus@times\relax}%
}
\newcommand{\o@plus@times}[2]{%
  \ooalign{$\m@th#1\oplus$\cr$\m@th#1\otimes$\cr}%
}
\DeclareRobustCommand{\bigbutthole}{%
  \mathbin{\mathpalette\bigo@plus@times\relax}%
}
\newcommand{\bigo@plus@times}[2]{%
  \ooalign{$\m@th#1\bigoplus$\cr$\m@th#1\bigotimes$\cr}%
}
\makeatother

\makeatletter
\newcommand{\bigzotimes}{%
  \DOTSB\mathop{\mathpalette\mattos@bigplus\relax}\slimits@
}
\newcommand\mattos@bigplus[2]{%
  \vcenter{\hbox{%
    \sbox\z@{$#1\sum$}%
    \resizebox{!}{0.9\dimexpr\ht\z@+\dp\z@}{\raisebox{\depth}{$\m@th#1\boxtimes$}}%
  }}%
  \vphantom{\sum}%
}
\makeatother

\newtheorem{theo}{Theorem}[section]
\newtheorem{eg}[theo]{Example}
\newtheorem{defn}[theo]{Definition}
\newtheorem{alg}[theo]{Algorithm}

\newcommand{\4}{\medskip \\}

\newcommand{\compc}{^\mathsf{c}}
\newcommand{\tranT}{^\mathsf{T}}

\newcommand{\RR}{\mathbb{R}}
\newcommand{\PP}{\mathbb{P}}

\newcommand{\NN}{\mathbb{N}}

\newcommand{\EE}{\mathbb{E}}

\newcommand{\TTT}{\mathcal{T}}

\newcommand{\PPP}{\mathcal{P}}

\newcommand{\MMM}{\mathcal{M}}

\renewcommand{\H}{\mathcal{H}}
\newcommand{\RRR}{\mathcal{R}}

\newcommand{\PPPP}{\mathscr{P}}

\newcommand{\vp}{\varphi}

\newcommand{\ztext}[1]{\; \text{ #1 } \; }

\newcommand{\ds}{\displaystyle}

\newcommand{\defeq}{\vcentcolon=}
\newcommand{\eqdef}{=\vcentcolon}

\renewcommand{\deg}{\mathsf{deg}}

\newcommand{\zve}{\varepsilon}

\newcommand{\za}{\alpha}

\newcommand{\zl}{\lambda}
\newcommand{\zo}{\omega}

\newcommand{\zt}{\theta}

\newcommand{\zd}{\delta}

\newcommand{\ve}{{\vec{e}}}
\newcommand{\pG}{\partial G}
\newcommand{\vE}{\vec{E}}
\newcommand{\fpv}{\mathsf{fpv}}
\newcommand{\len}{\mathsf{len}}
\newcommand{\CVR}{\text{CVR}}
\newcommand{\BAL}{\text{BAL}}
\newcommand{\vT}{\check{T}}
\newcommand{\vt}{\check{t}}
\newcommand{\vg}{\check{g}}

\newcommand{\hM}{\hat{M}}
\newcommand{\hMMM}{\hat{\mathcal{M}}}
\newcommand{\hmu}{\hat{\mu}}
\newcommand{\hk}{\hat{k}}
\newcommand{\hnu}{\hat{\nu}}
\newcommand{\vC}{\check{C}}
\newcommand{\zvt}{\vec{\theta}}
\newcommand{\zht}{\hat{\theta}}
\newcommand{\Var}{\text{Var}}
\newcommand{\zm}{\mathfrak{m}}

\newcommand{\first}{\mathsf{first}}

\newcommand{\hSE}{\widehat{\text{SE}}}

\newcommand{\pref}[3]{\hyperref[#1]{({\color{#3}\underline{#2}})}}
\newcommand{\cref}[3]{\hyperref[#1]{{\color{#3}\Circled{#2}}}}

\graphicspath{ {./Mathematica/} }

\begin{document}

\title{Graph-Based Audits\\for Meek Single Transferable Vote Elections}
\author{Edouard Heitzmann\textsuperscript{1}}
\date{\vspace{-4.0em}}
\begingroup
\renewcommand\thefootnote{\arabic{footnote}}
\maketitle

\footnotetext[1]{Department of Mathematics,\newline
	\mbox{}\hspace{16pt}University of Colorado, Boulder\newline
\mbox{}\hspace{16pt}\href{mailto:edouard.heitzmann@colorado.edu}{\nolinkurl{edouard.heitzmann@colorado.edu}}}
\endgroup

\stepcounter{footnote}

\begin{abstract}
	In the context of election security, a Risk-Limiting Audit (RLA) is a statistical framework that uses a minimal partial recount of the ballots to guarantee that the results of the election were correctly reported. A generalized RLA framework has remained elusive for algorithmic election rules such as the Single Transferable Vote (STV) rule, because of the dependence of these rules on the chronology of eliminations and elections leading to the outcome of the election. This paper proposes a new graph-based approach to audit these algorithmic election rules, by considering the space of all possible sequences of elections and eliminations. If we fix a subgraph of this universal space ahead of the audit, a sufficient strategy is to verify statistically that the true election sequence does not leave the fixed subgraph. This makes for a flexible framework to audit these elections in a chronology-agnostic way.
\end{abstract}
\section{Introduction.}

Single Transferable Vote (STV) methods are a family of algorithmic tabulation rules used to run multi-winner elections. They are currently used for political elections at the City Council level in Scotland, Northern Ireland, some US cities, and some regions of New Zealand. They are further used for state- or national-level elections in Australia, Malta, and Ireland.\4
There are many variants of STV, but all of them share the same underlying principle. They ask voters to fill out a full ranking over all candidates running for the $m$ seats to fill. During tabulation, candidates are eliminated or elected one-by-one. At each stage in this tabulation, each vote counts for its highest-ranked candidate still in the running (so-called `hopeful' candidates). Candidates are elected if they receive more than a pre-determined `quota' of votes, in which case some weight of the votes that got them quota is removed (or `exhausted') from the election. If no candidate makes quota, we eliminate the candidate with the fewest votes and transfer their votes onwards.\4 
STV has many desirable properties. It achieves proportional representation, in the sense that cohesive blocs of voters tend to be awarded a number of the $m$ seats proportional to their bloc's weight in the electorate. Unlike other proportional election rules, STV achieves this without requiring candidates to run with a partisan affiliation, which makes it suitable for non-partisan elections. As a multiwinner election method, it also has some natural insulation to gerrymandering. Finally, the turbulence of STV elections decreases the viability of strategic voting and incentivizes voters to represent their honest preferences.\4 
STV also has some drawbacks. Most of these drawbacks stem from its algorithmic nature, which makes it more complicated than other typical voting methods. This increases voter burden and makes election outcomes difficult to communicate, which can in turn lower trust in election outcomes. Algorithmic complexity also makes STV susceptible to pathologies like the so-called monotonicity, reinforcement, or no-show paradoxes -- see for example \textcite{graham-squire_paradoxical_2024}. These pathologies all point to the higher fragility of outcomes under STV: perturbing a small number of votes can have a big impact in an STV election. This is also what makes STV harder to audit than other election rules.\4 
A Risk Limiting Audit (RLA) of an election is, informally, a randomized partial recount of the physical ballots that gives some statistical guarantee that the outcome of the election was correctly reported. They are a way to make elections more secure while avoiding the high cost of doing a full hand recount of the election. Typically this is done by comparing the sampled ballots against their digital Cast Vote Records (CVRs), and checking that the discrepancy levels among sampled ballots would not be sufficient to overturn the reported margins of the election. \4 
The common wisdom is that audits are harder for STV because more rounds means more margins to audit. If we seek confidence about the exact chronology of the election -- every elimination and election from start to finish, as well as their timing -- there are more opportunities for decisions to hinge on razor-sharp margins, making the election unauditable. However, if we only seek confidence about the outcome of the election -- who won and who lost, without any details about the specific sequence that led to this outcome -- we find that much of this chronology can be overlooked, as it bears no impact on the outcome.\4 
To illustrate this point, consider District 1 of the 2024 Portland City Council election, which was a 3-seat STV contest\footnote{
The tabulation for this election is available \href{https://www.portland.gov/sites/default/files/2024/Portland-District-1-Certified-Abstract-Nov-2024.pdf}{\underline{here}}.}. In Round 6 of this election, candidates David Linn and Peggy Sue Owens were tied for last place with 1,141 first-place votes each, or 3.36\% of the ballots cast. If we wished to certify the exact chronology of this election, we would certainly have to perform a full recount to audit this margin of 0 votes. However, we find that which of these two candidates we choose to eliminate first has no bearing on the big picture of this election, since the second candidate is certain to be eliminated next (or, as we will find, before Round 9 of the election -- cf. Appendix \ref{sec:portland_d1}). But formalizing this insight mathematically is tricky.\4 
A further complication for these chronology-agnostic audits comes from the fact that in most variants of STV, the timing of a winner's election has an impact on subsequent Rounds of the election. If it does not matter whether we eliminate candidate A or B first, it does usually matter which of them is eliminated by the time C gets elected. This explains why a generalized STV auditing framework remains elusive, despite the success of the RAIRE framework for auditing single-winner STV contests in a chronology-agnostic way -- see \textcite{blom_raire_2019}. In developing this generalization, we show that not all variants of STV are equally suited to the task. We find that the Meek variant, being chronology-independent by definition, is more naturally auditable than the more commonly-used Weighted Inclusive Gregory Method (WIGM). That being said, the graph-based approach we develop for Meek STV can and should also be applied to more traditional variants of STV.\4 
The high-level picture of the graph-based approach we propose is to consider the space $\Omega$ of all possible sequences of elections and eliminations the election could follow. We pre-emptively carve out a subgraph $G\subseteq \Omega$, and we statistically verify that the true election path never leaves this subgraph $G$. As long as all the winner sets contained in the `bottom-line' of $G$ are the same as the reported one, this verifies the election result. We formalize the Experimental Design of this framework in Section \ref{sec:experimental_design}, after justifying our comparison between Meek and WIGM STV below.
\subsection{A Helpful Case Study.}\label{sec:case_study}
Meek and WIGM STV most often produce the same winner set given a fixed set of ballots -- they are, after all, variants of the same rule. To illustrate the difference between the two, we will focus on an example where this is not the case, to understand the root causes of their disagreement. We emphasize that such a disagreement is the exception, not the rule, for these two variants.\4 
One such example is the 2012 election in Ward 9 (Almond and Earn) of the Perth Kinross Councilmember election in Scotland. In this election, $N=3,689$ voters cast ballots to elect $m=3$ City Councilors using the WIGM variant of STV. In this variant, the quota of votes a candidate must receive to get elected is calculated using the Droop formula:
\[
	q = \left\lfloor\frac{N}{m+1}\right\rfloor +1 = 923.
\]
When a candidate's tally $T$ does exceed quota, the votes that got them there are re-weighted using a fractional transfer value $\tau$, which is calculated so as to make sure a quota's worth of votes is removed from the election, as follows:
\[
	\tau =\frac{T-q}{T}.
\]
For example, in Round 1 of our case study election, Alan Livingstone received $T=1,112$ first-place votes, so his transfer value was $\tau = 189/1,112 \approx 0.17$. Of the ballots that listed Livingstone as their first preference, 290 listed Alan Jack as their next preference, so Jack received a weight of around $290\cdot0.17 = 49.3$ ballots as a result of Livingstone's election. \4 
Crucially, in WIGM STV, this transfer process happens exactly once, and the transfer value is fixed for the rest of the election. In future rounds, if a ballot would transfer from another candidate to Livingstone, it instead goes straight to its next preference, or it is exhausted if none are specified, but Livingstone's transfer value is not recalculated. This is what we mean when we say WIGM STV is chronology-dependent: it matters which candidates were still in the running when Livingstone was elected.\4 
The full tabulation for the rest of the election is below:
\begin{table}[htbp]
  \centering
\begin{tabular}{|l|l|l|l|l|l|l|}
\hline
 & Round 1 & Round 2 & Round 3 & Round 4 & Round 5 & Round 6  \\
\hline
Alan Livingstone & 1112 & Elected &  Elected & Elected & Elected & Elected \\
\hline
Henry Anderson & 892 & 900.67 & {923.37} & Elected  & Elected & Elected  \\
\hline
Alan Jack & 444 & 493.29 & 596.93 & 596.96 & 770.62 & 986.97  \\
\hline
Wilma Lumsden & 628 & 640.41 & 670.96 & 671.23 & 756.43 & Eliminated  \\
\hline
Andrew Dundas & 369 & 381.24 & 450.7 & 450.73 & Eliminated & Eliminated  \\
\hline
George Hayton & 244 & 284.28 & Eliminated & Eliminated & Eliminated & Eliminated  \\
\hline
Quota & 923 & 923 & 923 & 923 & 923 & 923  \\
\hline
\end{tabular}
\caption{The tally breakdown in the 2012 Perth Kinross Ward 9 election, as it occurred in reality.}
\end{table}\\
This election is peculiar in two ways: firstly, Henry Anderson is elected by a razor sharp margin of $0.37$ votes in Round 3, and secondly, Wilma Lumsden narrowly loses to Alan Jack in Round 5 by a margin of $14.19$ votes. As it turns out, Anderson and Lumsden were both candidates from the Scottish National Party (SNP), meaning that many voters who listed one candidate also listed the other. Of Anderson's surplus of $0.37$ votes, $0.27$ transfers to Lumsden -- around 73\%!\4
In Round 4, Andrew Dundas is eliminated. Of the votes he transfers onwards, $85.2$ transfer to Lumsden, and $173.65$ go to Jack. Of the remaining weight of $191.87$ votes in Dundas' pile, an additional $39.51$ would have listed Anderson as their next preference, but do not mention Lumsden. According to the WIGM rules of STV, those votes are exhausted from the election, since Anderson is already elected, and his transfer value does not change. \4 
This makes the election very unstable -- if Anderson's tally had $0.37$ fewer votes in Round 3, those $39.51$ votes would have transferred through Anderson before exhausting (along with an additional $35.34$ votes that listed Anderson and then Lumsden), pushing Anderson's transfer value high enough that Lumsden would now edge out Jack in Round 5 of the election. This can be realized in a number of ways:
\begin{enumerate}
	\item As a monotonicity paradox: by replacing one vote 
		\[
			(\text{Henry Anderson},\text{Alan Jack})\rightarrow (\text{Alan Jack}, \text{Henry Anderson}),
		\]
		we would be increasing the strength of the profile's support for Jack in a way that loses him the election.
	\item As a failure of the Independence of Irrelevant Alternatives axiom: by replacing a single bullet vote 
		\[
			(\text{Henry Anderson}) \rightarrow (\text{George Hayton}),
		\]
		we would not be changing any of the profile's comparisons between Lumsden and Jack, yet the outcome's comparison between these candidates would change (Lumsden would now beat Jack).
	\item As a no-show paradox: if a single voter 
		\[
			(\text{Henry Anderson}, \text{Wilma Lumsden})
		\]
		had decided not to cast her vote in the election, Lumsden (one of her preferred candidates) would have been elected. Indeed, up to 18 such voters (out of 646 recorded) could have abstained while still preserving that outcome.
\end{enumerate}

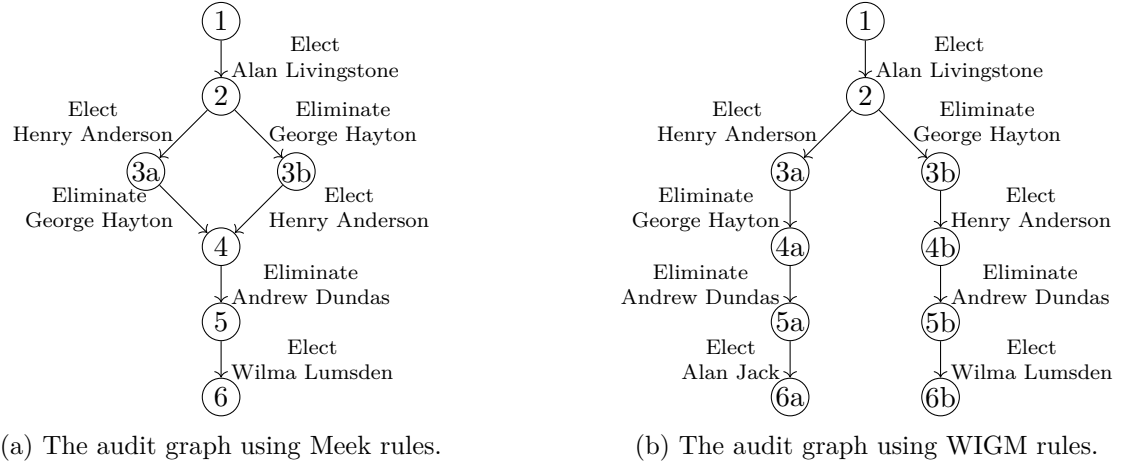
\begin{figure}[h]
  \centering

  \begin{subfigure}[b]{0.48\textwidth}
    \centering
    \begin{tikzpicture}[scale=1]
      \tikzset{
        graphnode/.style={circle, draw, minimum size=0.5cm, inner sep=0pt},
        edgelabel/.style={font=\scriptsize, align=center}
      }

      \node[graphnode] (n1) at (0,4) {1};
      \node[graphnode] (n2) at (0,3) {2};
      \node[graphnode] (n3) at (-1,2) {3a};
      \node[graphnode] (n4) at (1,2) {3b};
      \node[graphnode] (n5) at (0,1) {4};
      \node[graphnode] (n6) at (0,0) {5};
      \node[graphnode] (n7) at (0,-1) {6};

      \draw[->]  (n1) -- node[edgelabel, right]{Elect\\Alan Livingstone} (n2);
      \draw[->]  (n2) -- node[edgelabel, left]{Elect\\Henry Anderson\\} (n3);
      \draw[->]  (n2) -- node[edgelabel, right]{Eliminate\\George Hayton\\} (n4);
      \draw[->]  (n3) -- node[edgelabel, left]{Eliminate\\George Hayton} (n5);
      \draw[->]  (n4) -- node[edgelabel, right]{Elect\\Henry Anderson} (n5);
      \draw[->]  (n5) -- node[edgelabel, right]{Eliminate\\Andrew Dundas} (n6);
      \draw[->]  (n6) -- node[edgelabel, right]{Elect\\Wilma Lumsden} (n7);
    \end{tikzpicture}
    \caption{The audit graph using Meek rules.}
    \label{fig:ward9-meek}
  \end{subfigure}
  \hfill
  \begin{subfigure}[b]{0.48\textwidth}
    \centering
    \begin{tikzpicture}[scale=1]
      \tikzset{
        graphnode/.style={circle, draw, minimum size=0.5cm, inner sep=0pt},
        edgelabel/.style={font=\scriptsize, align=center}
      }

      \node[graphnode] (n1) at (0,4) {1};
      \node[graphnode] (n2) at (0,3) {2};
      \node[graphnode] (n3) at (-1,2) {3a};
      \node[graphnode] (n4) at (1,2) {3b};
      \node[graphnode] (n5) at (-1,1) {4a};
      \node[graphnode] (n6) at (1,1) {4b};
      \node[graphnode] (n7) at (-1,0) {5a};
      \node[graphnode] (n8) at (1,0) {5b};
      \node[graphnode] (n9) at (-1,-1) {6a};
      \node[graphnode] (n10) at (1,-1) {6b};

      \draw[->]  (n1) -- node[edgelabel, right]{Elect\\Alan Livingstone} (n2);
      \draw[->]  (n2) -- node[edgelabel, left]{Elect\\Henry Anderson\\} (n3);
      \draw[->]  (n2) -- node[edgelabel, right]{Eliminate\\George Hayton\\} (n4);
      \draw[->]  (n3) -- node[edgelabel, left]{Eliminate\\George Hayton} (n5);
      \draw[->]  (n4) -- node[edgelabel, right]{Elect\\Henry Anderson} (n6);
      \draw[->]  (n5) -- node[edgelabel, left]{Eliminate\\Andrew Dundas} (n7);
      \draw[->]  (n6) -- node[edgelabel, right]{Eliminate\\Andrew Dundas} (n8);
      \draw[->]  (n7) -- node[edgelabel, left]{Elect\\Alan Jack} (n9);
      \draw[->]  (n8) -- node[edgelabel, right]{Elect\\Wilma Lumsden} (n10);
    \end{tikzpicture}
    \caption{The audit graph using WIGM rules.}
    \label{fig:ward9-wigm}
  \end{subfigure}

  \caption{The audit graphs for an auditable margin of $40$ votes in Ward 9.}
  \label{fig:ward9-audittrees}
\end{figure}

The change caused by these three perturbations is the same -- even though fewer votes transfer from Dundas to the Anderson-Lumsden slate than to Jack, the ``lost'' weight of $39.51$ gets to transfer into the slate with this new elimination order; and even though these votes only listed Anderson and not Lumsden (indicating a non-solid coalition), this difference is enough to overcome the $14.19$ vote margin that Lumsden was losing by.\4 
It is normatively nebulous to say that Lumsden ``deserved'' to win more than Jack in this election. Both candidates were competitive, and the election was sharp -- in most of the alternate elimination orders that lead to Lumsden winning under WIGM rules, she only beats Jack by a margin of $\sim 9.5$ votes. But it is fair to say that the timing of his election was cruel to Anderson in this case: through no fault of his own, Anderson was elected with a minuscule surplus, which disenfranchised over $39.51$ voters who would have ranked him later on, and diminished the next preferences of the votes that got him quota.\4 
The rules of Meek STV were designed to address exactly this scenario. At a high level, Meek changes two rules compared to WIGM: it allows winners to receive votes even after the Round of their election, in which case they increase their transfer value, and it re-calculates quota every round, lowering it when votes are exhausted because they list no further preference in the election. Because of the former difference, the $39.51$ votes we mentioned still get added to Anderson's pile, even after he has attained quota, and Anderson gets to increase his transfer value. Even though the new $39.51$ votes do not list Lumsden as their next preference, a rising tide lifts all boats, and the rest of Anderson's pile gets to transfer with an increased weight to Lumsden. The result of re-tabulating the same election using the Meek rule is shown in Table \ref{table:meek_ward9}. 
\begin{table}[H]
  \centering
\begin{tabular}{|l|l|l|l|l|l|}
\hline
 & Round 1 & Round 2 & Round 3 & Round 4 & Round 5  \\
\hline

Alan Livingstone & 1112.00 \checkmark& 904.07 \checkmark & 885.12 \checkmark & 884.13 \checkmark & 833.50 \checkmark \\
\hline
Henry Anderson & 892.00 & 901.54 & 926.30 \checkmark & 884.13 \checkmark & 833.50 \checkmark \\
\hline
Alan Jack & 444.00 & 498.23 & 608.11 & 612.87 & 810.07  \\
\hline
Wilma Lumsden & 628.00 & 641.65 & 674.34 & 705.24 & 856.94 \checkmark \\
\hline
Andrew Dundas & 369.00 & 382.46 & 446.60 & 450.16 & Eliminated  \\
\hline
George Hayton & 244.00 & {288.32} & Eliminated & Eliminated & Eliminated  \\
\hline
Quota & 922.25 & 904.07 & 885.12 & 884.13 & 833.50  \\
\hline

\end{tabular}
\caption{The Ward 9 election tally breakdown under Meek STV.}\label{table:meek_ward9}
\end{table}
A happy side-effect is that the Meek election is a lot more resilient to perturbation: if we allow perturbations of less than $40$ votes' worth of weight, the only real point of uncertainty is whether Anderson should have been elected in Round 2 or Round 3 -- but whichever round the election occurs in, the other round will result in the elimination of George Hayton, such that both branches converge back to the same election state before Round $4$. We begin to formalize this notion of stability with the construction of `audit graphs' in Figure \ref{fig:ward9-audittrees}.
\4
Starting with the `election state' $(H, W) = (C, \varnothing)$ in vertex \Circled{1}, we draw an edge to every election state $(H,W)$ we might plausibly encounter in the next round of the election if the recorded tallies (the so-called `Cast Vote Records,' or CVRs) were off by at most $40$ votes, repeating this process until we end up at nodes with complete winner sets. If all vertices in the last layer of the resulting graph $G$ have the same winner set, then a valid strategy to audit the election would be to reject the global null hypothesis $\H_\star = $ ``The true election order leaves the graph $G$.'' If we can reject $\H_\star$, we might not know exactly which path the true election takes from node $\Circled{1}$ to the last layer of the graph, but we know that every such path leads to the same winner set, confirming the result of the election.\4
The audit graphs in Figure \ref{fig:ward9-audittrees} demonstrate that the Meek rule is more stable than WIGM in this case. Despite having the same sets of elected winners and hopeful candidates, the election states represented by nodes $\Circled{4a}$ and $\Circled{4b}$ in Figure \ref{fig:ward9-wigm} must be considered distinct, since Anderson has a lower transfer value in the former than the latter. By contrast, the edges leaving nodes $\Circled{3a}$ and $\Circled{3b}$ in Figure \ref{fig:ward9-meek} can truly be considered to lead to the same election state, since whichever path we take to node $\Circled{4}$ will result in the same calibrated transfer values.\4
Hence this is an example of an election which is unauditable under WIGM rules (since $0.37$ of a vote can change the outcome of the election, a full recount will surely be necessary), but which is auditable under Meek rules, where the sharpest margin to audit is the $46.87$ vote margin between Lumsden and Jack in the last round of the election. Of course, there are other profiles where Meek will be unauditable for other reasons (and WIGM might be auditable in some of these profiles); but this example illustrates that uncertain election chronology poses a problem in WIGM audits, whereas it does not in Meek.
\subsection{Mathematical Formalism.}
The STV rules are algorithmic processes to determine a committee of $m$ winners from a set of $C = [M]= \{0,1,\ldots, M-1\}$ candidates, by consulting a \textbf{preference profile} $P$ containing the ranked choice votes (or just \textbf{rankings}) of a set of voters $[N] = \{0,1,\ldots, N-1\}$. We define these mathematically as follows:
\begin{defn} (Rankings, Profiles.) \label{defn:rankings} 
            \begin{itemize}
	\item Given $L\in \NN$ and a subset of ``hopeful'' candidates $H\subseteq C$, an $\boldsymbol{L}\textbf{-ranking}$ of candidates $H$ is an injective map
	\[
		r\colon [L]\hookrightarrow H.
	\]
	We allow $L=0$, in which case $[L]=\varnothing$, and the only ranking $r:\varnothing \to H$ is the empty ranking $r=\varnothing$; this corresponds to a vote which indicated no preferences in $H$.
	\item We denote the set of all rankings of candidates $H\subseteq C$ by $R_H$:
		\[
			R_H \defeq \{r\colon [L]\hookrightarrow H \mid 0\leq L\leq |H|\}.
		\]
	\item For $H\subseteq C$, the \textbf{length map} $\len\colon R_H\to [|H|+1]$ maps each $r$ to the integer $L$ such that $r$ is an $L-$ranking; formally,
		\[
			\len(r) = |r|.
		\]
	\item A \textbf{profile} $P$ on candidates $H$ can be thought of as a multiset of rankings, or as a map $P\colon R_H\to \NN$, or as a point in an $|R_H|-$dimensional simplex. In the context of auditing, it will be useful for these profiles to be ordered, so we formally define a profile on candidates $H\subseteq C$ as a map
		\[
			P\colon [N] \to R_H.
		\]
	\item When the number of voters $N>0$ is clear, we denote the set of all profiles on candidates $H$ by $\PPPP_H = \{P\colon [N]\to R_H\}$, or just $\PPPP$ when $H$ is also clear.
\end{itemize}
Given a ranking $r\in R_C$, we think of $r(0)$ as the most preferred candidate of $r$, then $r(1)$ as its second-most preferred, etc.
\end{defn}
In this formalism, the STV rules can be thought of as maps $\PPPP_C\to\{W\subseteq C : |W|=m\}$. These rules are algorithmic: starting from the full candidate set $C$, it iteratively elects or eliminates one candidate at a time, until $m$ candidates have been elected. One advantage of the Meek rule is that it can be described using elimination rounds only -- election rounds are not necessary. In this framework, given a fixed profile $P$, we can view the Meek rule as a monotone decreasing map 
\[
	\RRR_P\colon \{H\subseteq C : |H| >m\} \to \PPP(C)\ztext{such that} \RRR_P(H)\subseteq H \ztext{and}|\RRR_P(H)| = |H|-1,
\]
where $\PPP(C)$ is the power set of $C$. Then the winner set under a profile $P$ on candidates $C=[M]$ is defined as 
\[
	W = \RRR_{P}^{M-m}(C).
\]
Before we can describe the STV rules mathematically, we need to establish how to project a profile of rankings onto a set of remaining hopeful candidates $H\subseteq C$. This works as one might expect: to project a ranking $r\in R_H$ onto $H'$, we `erase' the positions in $r$ that list candidates not in $H'$, and `shift' all the remaining positions to fill any empty positions.
\begin{defn} (Projection onto hopeful candidates, first-place vote map.)\4
	For fixed sets of hopeful candidates $H'\subseteq H\subseteq C$, given an $L$-ranking $r\in R_H$ such that $|r([L])\cap H'| = L'$, we define the \textbf{shift map} of $r$ as the unique increasing injection
	\[
		J_r\colon [L'] \hookrightarrow [L]\ztext{such that} r \circ J_r([L'])\subseteq H'.
	\]
	In the same setting, we define the \textbf{ranking projection map} $\pi_{H'}\colon R_H\to R_{H'}$ by
	\[
		\pi_{H'}(r) = r\circ J_r.
	\]
	Further, the \textbf{projection onto $\boldsymbol{H'}$ of a profile} $P\in \PPPP_H$ is defined as 
	\[
		\pi_{H'}\circ P\in \PPPP_{H'}.
	\]
	Finally, it will be useful to define the \textbf{first place vote map} $\fpv_H\colon R_C\to H\cup\{-1\}$ as 
	\[
		\fpv_H(r) = \begin{cases}
			\big(\pi_H(r)\big)(0) & \ztext{if} \pi_H(r) \neq \varnothing \\ 
		-1 & \ztext{if} \pi_H(r) = \varnothing. \end{cases}
	\]
\end{defn}
We are presently equipped to describe the STV rules. \4 
The WIGM (again, Weighted Inclusive Gregory Method) Rule has seen wide enough use to be considered a piece of folklore; its genealogy seems to go back to 1880, although the first record of it appears in \textcite{humphreys_proportional_1911}, where it is attributed to one `Mr. J. B. Gregory of Melbourne.' WIGM is notorious for having as many variants as there are governments that use it -- some calculate transfer values differently, or handle simultaneous elections differently, or calculate quota differently, etc. We formalize the version of WIGM we will henceforth refer to below.
\begin{alg} (The WIGM Rule.)\label{defn:WIGM}\4
	Given a profile $P\in\PPPP_C$ of $N$ voters with the additional constraint that $P(i)\neq\varnothing\;\forall\; 0\leq i\leq N-1$ (i.e. each voter indicates a preference among the initial candidates $C$), the \textbf{WIGM rule for $\boldsymbol{m}$ seats} iterates through the following steps to find a winner set $W\subseteq C$:
	\begin{itemize}
		\item[\Circled{0}] Calculate the so-called \textbf{Droop quota} $q$ for this election as 
		    \[
			q = \left\lfloor\frac{N}{m+1}\right\rfloor+1.
		    \]
		    Further initialize the set of winners as $W=\varnothing$, the set of hopeful candidates as $H= C$, and define the initial \textbf{voter weight map} $w_0\colon [N]\to [0,1]$ as the constant map $w_0(i)=1\;\forall\; i\in[N]$. Then, start iterating the following steps with $j=0$.
	    \item[\Circled{1}] If $|H\cup W|=m$, the algorithm is complete, and the winner set is $H\cup W$.
    	\item[\Circled{2}] For each $c\in H$, define $S_c = \{i\in[N]:\fpv_H(P(i)) = c\}$. Calculate this round's \textbf{tallies} $T_j\colon H\to \RR^+$ as 
		\[
			    T_j(c) = \sum_{i\in S_c} w_j(i).
		\]
	\item[\Circled{3}] If any of the hopeful tallies are at or above quota, this is an election round: go to step \Circled{3a}. Otherwise, this is an elimination round: go to step \Circled{3b}.
		\begin{itemize}
	\item[\Circled{3a}] let $c$ be the candidate with the highest\footnotemark[2]{} tally above quota, i.e. $c$ is such that 
		\[
			T_j(c)\geq T_j(i)\;\forall \; i\in H \ztext{and} T_j(c)\geq q.
		\]
		Calculate the \textbf{transfer value} $\tau_c$ of this winner as 
		\[
			\tau_c = \frac{T_j(c)-q}{T_j(c)}.
		\]
		Use this transfer value to update the voter weight map for the next round:
		\[
			w_{j+1}(i)=\begin{cases}
				\tau_c\cdot w_j(i) & \ztext{if} i\in S_c; \\ 
				w_j(i) &\ztext{otherwise.}
			\end{cases}
		\]
		Finally, remove $c$ from $H$ and add her to $W$, then increment $j$ and return to step \Circled{1}.
	\item[\Circled{3b}] let $\ell$ be the hopeful candidate with the lowest\footnotemark[3]{} tally, i.e. $\ell\in H$ is such that
		\[
			T_j(\ell)\leq T_j(i) \; \forall\; i\in H.
		\]
		Finally remove $\ell$ from $H$, increment $j$, and return to step \Circled{1}, using the same voter weight map $w_{j+1}=w_j.$
	\end{itemize}
	\end{itemize}	
	\footnotetext[3]{If there are multiple such $\ell$, break the tie in a reproducible way, e.g. by picking the smallest such index (a so-called `lexicographic' tiebreak).}\stepcounter{footnote} 
\end{alg}
Before we formally define the Meek Rule, let us highlight its main differences compared to WIGM, to draw a contrast:
\begin{itemize}
    \item Meek uses so-called `keep factors' $k$ instead of transfer values $\tau$. If nothing else was different about Meek, we could think of these as just the complement of the transfer values: $k = 1-\tau$.
    \item In Meek, winners are allowed to receive transfers, even after the round of their election. When this happens, the winner re-calculates her keep factor from scratch, lowering it to adjust her tally (including the new votes she received) at exactly the current value of quota.
    \item We also re-calculate quota every round in Meek -- unlike WIGM, this value is not fixed at the start of the election. When ballots exhaust from the election (i.e. all preferences on the ballot are either elected or eliminated, but the vote still carries the weight), we treat the remaining weight of the ballot as though it did not figure in the total ballot weight used to calculate Droop quota, hence quota lowers.
    \item Because of situations where multiple seated winners transfer ballots to each other, or situations where decreasing the value of quota causes new votes to exhaust, the first round of keep factor calculations often does not calibrate the tallies of all winners at the value of current quota -- quota either lowers after new votes are exhausted from winner tallies, or winner tallies increase above quota after receiving votes from other winners. Because of this, we use a recursive process which re-calculates keep factors iteratively until winner tallies have `converged' to quota in a well-defined sense.
\end{itemize}
\begin{alg} (The Meek rule.) \label{defn:meek_rule}\4 
	Given a profile $P\in \PPPP_C$ of $N$ voters, a \textbf{minimum surplus} $\zve$, and a \textbf{tolerance} $\omega$ (usually both set to $\zve,\zo = 10^{-6}$), the \textbf{Meek rule for $\boldsymbol{m}$ seats} iterates through the following steps to find a winner set $W\subseteq C$:
	\begin{itemize}
		\item[\Circled{0}] Start by setting $H=C$.
		\item[\Circled{1}] Let $P_H=\pi_H\circ P$.
		\item[\Circled{2}] Initialize the first set of \textbf{keep factors} $k_0\colon H\to(0,1]$ as the constant map $k_0(c)=1, \:\forall \:c\in H$. Then, iterate through the following steps (starting with $j=0$) to calibrate the keep factors:
			\begin{itemize}
				\item[\Circled{2a}] Let $V$ be the $|H|$-dimensional real vector space with basis $\{\ve_c\}_{c\in H}$. Define this iteration's \textbf{weight vector map} $w_j\colon R_H\to V$ by 
					\[
						w_j(r) = \sum_{i=0}^{\len(r)-1} k_j(r(i)) \left(\prod_{l = 0}^{i-1}\big(1-k_j(r(l)\big)\right)\ve_{r(i)}.
					\]
					N.b.: depending on the keep factors $k_j$ and the ranking $r$, the total weight $\sum_{i\in H} \big(w(r)\big)(i)$ may be smaller than $1$! Some or all of the weight of $r$ might be \textbf{exhausted}; e.g., if $r=\varnothing$ has no preferences, then $w(r) = \vec{0}$.
				\item[\Circled{2b}] Define this iteration's \textbf{tallies} $T_j\in V$ as 
					\[
						T_j = \sum_{i=0}^{N-1} w_j\circ P(i).
					\]
				\item[\Circled{2c}] Calculate this iteration's quota $q_j\in \RR$ as 
					\[
						q_j = \frac{\sum_{i\in H} T_j(i)}{m+1}+\zve.
					\]
				\item[\Circled{2d}] Break out of this loop if all of the tallies in $T_j$ are less than tolerance $\omega$ above quota, i.e. if 
					\[
						\max_{i\in H}(T_j(i)-q_j) <\omega.
					\]
					In this case, go to step \Circled{3}; otherwise, proceed with step \Circled{2e}.
				\item[\Circled{2e}] Calculate the next iteration's keep factors $k_{j+1}\colon H\to (0,1]$ as follows:
					\[
						k_{j+1}(i) = k_j(i)\cdot\min\left(1, \frac{q_j}{T_j(i)}\right).
					\]
					Then, increment $j$ and return to step \Circled{2a}.
			\end{itemize}
		\item[\Circled{3}] Let $\ell\in H$ be the index of the {lowest\footnotemark} tally in the last iteration of tallies $T_j$ from the loop in \Circled{2}; i.e. $\ell$ is such that
			\[
				T_j(\ell) \leq T_j(i)\;\forall\;i\in H.
			\]
		Define $H' = \RRR_P(H) = H\backslash\{\ell\}$. If $|H'|=m$, the algorithm is complete and $W= H'$ are the winners of the election. Otherwise, return to step \Circled{1}, setting $H=H'$.
	\end{itemize}	
\footnotetext{Again, breaking ties in a consistent way, e.g. lexicographically.}
It is worth noting that when Meek is used in actual elections, it \emph{is} typical to distinguish candidates $c$ as winners the first time their keep factor $k(c)$ dips below $1$, unlike the version of the algorithm we describe. In such an election round it is common to break out of the loop in step \Circled{2} as soon as a new candidate is detected above quota. This has the effect of making the tallies $T(i)$ look like they weren't fully calibrated at quota (since we ended the iteration before the keep factors converged); but mathematically speaking this will always produce the same effect as our version of the algorithm, because in the next round of the election we would begin calibrating the keep factors from scratch at $1$.\4 
The Meek rule was originally outlined in \textcite{meek_nouvelle_1969}, and its algorithmic properties -- in particular, the convergence of step \Circled{2} -- were further studied in \textcite{hill_algorithm_1987}.

\end{alg}
The last notion we should formalize is that of a \textbf{Ballot-Comparison Risk-Limiting Audit (RLA)}.\4
In the real-world setting, elections are often tabulated by applying the election rule to a digital \textbf{Cast Vote Record (CVR)} of the paper ballots. We consider this profile to be a noisy approximation of the true preference profile, which is recorded in the paper ballots cast by the voters in the election; noise can be introduced by scanning or transcription errors, by interference, or by miscommunication of intent. Usually, doing a full recount of this true profile is expensive, so it is preferable instead to construct a series of statistical assertions about it that can be verified by looking at a small sample of paper ballots to confirm the result of the election.
\begin{defn} (Risk-Limiting Audits.)\4  
	We define an \textbf{election} as a pair of profiles on candidates $C$ with the same number $N$ of voters, $\BAL, \CVR\in\PPPP$. The \textbf{noise level} of this election is defined as
			\[
				\eta = \frac{|\{0\leq i<N : \BAL(i) \neq \CVR(i)\}|}{N}.
			\]	
		Given an $m$-seat voting rule $\RRR\colon \PPPP\to \{W\subseteq C:|W|=m\}$, the \textbf{recorded winner set} of the election is defined as $W_\star = \RRR(\CVR)$. A \textbf{Risk-Limiting Audit (RLA)} for this election is statistical framework to verify $\RRR(\BAL)= W_\star$ by sampling as few entries from $\BAL$ as possible. This usually takes the form of a \textbf{global null hypothesis $\boldsymbol{\H_\star}$} about $\BAL$ whose rejection would guarantee that $\RRR(\BAL)= W_\star$.  \4 
	Given a fixed sample size $n \leq N$, a \textbf{Ballot-Comparison Compiler} for this global null is a rejection rule $R_\star$ for $\H_\star$ that takes for input a set of $n$ ranking pairs (thought of as a sample of ballot-CVR pairs that we are `comparing' against each other to look for discrepancies):
			\[
				R_\star \subseteq (R_C \times R_C)^n.
			\]
	In order to decide whether we should reject the null $\H_\star$, we sample $n$ pairs of ballot-CVR pairs without replacement and we check if this sample falls within our rejection region $R_\star$. The \textbf{risk level} of an RLA with global null $\H_\star$ and rejection rule $R_\star$ is a bound on the probability $\za$ of making a Type I error:
	\[
		\za \geq \PP(R_\star \mid \H_\star).
	\]
\end{defn}
The concept of RLAs has been around since at least 2009 -- see \textcite{lindeman_gentle_2012} for a standard reference. The interest in auditing algorithmic election rules is a more recent development: see \textcite{blom_3_2025} for the current state of the art approach. We borrow the word `compiler' from this literature, although we define it in a slightly different way; rather than a rejection rule, compilers would traditionally be defined as a sequential $p$-value for the global null $\H_\star$ (or rather, for each of the local nulls that compose it). These $p$-values are usually given by a submartingale; in our framework, the rejection rules will come from Wald confidence intervals constructed via the delta method.
\section{Experimental Design.} 

\subsection{Graph-Based Audits.}\label{sec:experimental_design}
Before we state our main result, we need to borrow a definition from the Algorithmic Graph Theory literature -- cf. \textcite{fiat_competitive_1991} for instance:
\begin{defn} (Layered Graphs.) \4
A \textbf{directed layered graph $\boldsymbol{G= (V, \vE)}$} is a directed, rooted, and connected graph whose vertices $V$ are partitioned into layers $L_0\sqcup L_1\sqcup \ldots \sqcup L_M$, such that $L_0 =\{C\}$ contains only the root of $G$, and all the edges in $\vE$ connect only vertices from one layer to the next:
	\[
		\vE \subseteq \bigcup_{i=0}^{M-1}L_i\times L_{i+1}.
	\]
\end{defn}
The following result is not a novelty in the world of hypothesis-testing -- it is a specialized version of an \textbf{intersection-union} test, seen for example in Theorem 1 of \textcite{berger_multiparameter_1982}. Still, this result bears emphasis because of how magical it appears to be: when the global null $\H_\star$ implies a disjunction of local nulls $\H_\ve$, a sufficient $\za$-level test is to reject every $\H_\ve$ at level $\za$. Fortunately for our graph-based auditing approach, where there are often hundreds of local nulls, there is no need to budget the risk level over each local null. 

\begin{customnumbered}[($\boldsymbol{\za}$-level test for $\boldsymbol{\H_\star}$)] \label{prop:design}
    Let $\Omega$ be a directed layered graph with root $C$, let $G\subseteq \Omega$ be a connected subgraph containing $C$, and define $\partial G \defeq G\times G^{\mathsf{c}}\cap \vec{E}_{\Omega}$. Consider a process which moves from $C$ to the last layer of $\Omega$ by following edges according to a rule $\RRR_P\colon\Omega\to\Omega$ determined by a parameter $P$ drawn from a parameter space $\PPPP$. Once $P$ is fixed, the process deterministically moves from each vertex $u\in\Omega$ to $\RRR_P(u)\in N(u)$, drawing out a path from $C$ to the last layer of $\Omega$.\4 
    Let $\H_\star$ be the null hypothesis ``The process leaves the subgraph $G$.'' For each $\ve = (u,v)\in \pG$, let $\H_\ve$ be the null hypothesis ``$\RRR_P(u)=v$.'' Given a set of $\za$-level rejection rules $\{R_\ve\subseteq \PPPP\}_{\ve\in\pG}$ for the $\H_\ve$, a valid $\za$-level rule for $\H_\star$ is 
    \[
        R_\star \defeq \bigcap_{\ve\in\pG} R_\ve.
    \]
    In other words, reject $\H_\star$ at level $\za$ if and only if we reject every $\H_\ve$ at level $\za$.
\end{customnumbered}
\textbf{Proof.} Let $\PPPP_\star\subseteq \PPPP$ be the subset of the parameter space where $H_\star$ is true. For each $\ve \in \pG$, let $\PPPP_\ve\subseteq \PPPP$ be the subset of the parameter space where $\H_\ve$ is true. \4 
Suppose that $P_0\in \PPPP_\star$, i.e. the process defined by $P_0$ leaves $G$. Then it must at some point get to a vertex $u_0\in G$ such that $u_0\defeq\RRR_{P_0}(v_0)\notin G$, hence in particular $H_{(v_0,u_0)}$ is true, and $P_0\in \PPPP_{(u_0,v_0)}$. Hence for such a $P_0$ we have 
\[
	\PP_{P_0}(R_\star)\leq \sup_{P\in \PPPP_{(u_0, v_0)}}\PP_{P}(R_{(u_0,v_0)}) \leq \za.
\]
Since this holds true for any $P_0\in \PPPP_\star$, we see \vspace{-6pt}
\[
	\sup_{P\in\PPPP_\star}\PP_P(R_\star) \leq \za. \tag*{\qed}
\]
\begin{eg}\label{eg:P_6}
    Let $\PPPP=(\{0,1\}^6)^{1000}$ be the space of all samples of size $1000$ from the 6-dimensional hypercube, and consider the graph $\Omega$ given below.\4
    \begin{minipage}{0.63\textwidth}
	Let the subgraph $G$ consist of nodes 1a, 2a, 3a, and 4a. Then the edges in $\partial G$ are drawn with dashed lines. To traverse the graph $\Omega$, let $P_i$ represent the number of $1$s in the $i^{\text{th}}$ components of the elements in $P$. We define 
	\[
		\RRR_P(\text{1a}) = \begin{cases}
			\text{2b} & \text{if} \; P_1 \geq 200; \\
			\text{2c} & \text{if} \; P_1 < 200 \ztext{and} P_2 \geq 200;\\
			\text{2a} & \text{otherwise.}
		\end{cases}
	\]
	The rules for the other nodes $(i)a$ are similar -- we follow the edge labelled $i$ if $P_i\geq 200$ and $P_{i-1}< 200$, otherwise we move to node $(i+1)a$.
\end{minipage}
\hfill
\begin{minipage}{0.34\textwidth}
	\begin{center}
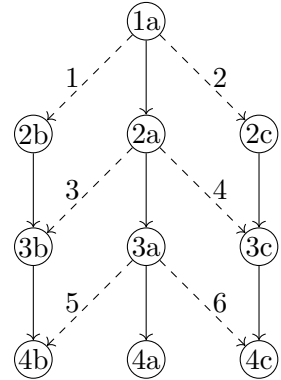

	\begin{tikzpicture}[scale=1.5]
	  \tikzstyle{graphnode}=[circle, draw, minimum size=0.5cm, inner sep=0pt]
	  \node[graphnode] (a1) at (0,3) {1a};
	  \node[graphnode] (a2) at (0,2) {2a};
	  \node[graphnode] (a3) at (0,1) {3a};
	  \node[graphnode] (a4) at (0,0) {4a};
	  \node[graphnode] (b1) at (1,2) {2c};
	  \node[graphnode] (b2) at (1,1) {3c};
	  \node[graphnode] (b3) at (1,0) {4c};
	  \node[graphnode] (c1) at (-1,2) {2b};
	  \node[graphnode] (c2) at (-1,1) {3b};
	  \node[graphnode] (c3) at (-1,0) {4b};

	  \draw[->] (a1) --  (a2);
	  \draw[->] (a1) -- node[right]{2} (b1) [dashed];
	  \draw[->] (a1) -- node[left]{1} (c1) [dashed];
	  \draw[->] (a2) --  (a3);
	  \draw[->] (a2) -- node[right]{4} (b2) [dashed];
	  \draw[->] (a2) -- node[left]{3} (c2) [dashed];
	  \draw[->] (a3) --  (a4);
	  \draw[->] (a3) -- node[right]{6} (b3) [dashed];
	  \draw[->] (a3) -- node[left]{5} (c3) [dashed];
	    
	  \draw[->] (b1) --  (b2);
	  \draw[->] (b2) --  (b3);
	  
	  \draw[->] (c1) --  (c2);
	  \draw[->] (c2) --  (c3);
	\end{tikzpicture}
\captionof{figure}{\centering The Universal Graph $\Omega$ \\ 
in Example \ref{eg:P_6}.}
\end{center}
\end{minipage} \mbox{}\vspace{6pt}\\ 
Here our global null is $\H_\star=$``The process leaves $G$,'' which is equivalent to ``The process does not end at vertex 4a.'' This global null implies a disjunction of local nulls $H_i$=``The process follows the edge labelled $i$.''\4
To test the hypotheses $\H_i$ for $1\leq i \leq 6$, we might look at the first 100 samples in $P$ and count the number of $1$s we see in each component, then use them to build a one-sided confidence interval for $P_i$. To reject $\H_i$, it is sufficient to require $\PP(P_i\geq200)<\za$. \4 
So let $\pi_i$ represent the number of $1$s in the $i^{\text{th}}$ component of the first 100 samples in $P$. Picking $\za=0.05$, the hypergeometric distribution for $k=\pi_i$ successes out of $n=100$ draws from a population of size $N=1000$ with $K=P_i$ total successes tells us that a sufficient rule is to reject $H_i$ when $\pi_i<14$:
\[
	R_i = \{P\in\PPPP\colon \pi_i(P)<14\}, \ztext{because} \PP(\pi_i<14\colon P_i = 200) \approx 0.039 < \za.
\]
Therefore our global rule is to reject $H_\star$ (and conclude that the process terminates at vertex 4a) when the total number of $i^{\text{th}}$ component $1$s in the first 100 elements of $P$ is below $14$ for each $1\leq i \leq 6$. The first row of Table \ref{table:examples_3} shows that this rule is very conservative: when $P$ is such that $P_i=200$ for all $i$ -- exactly on the boundary between $\H_\star$ being true or false -- this rule never rejects $\H_\star$ out of $50,000$ simulated trials.
\end{eg}

Perhaps one of the better ways to illustrate Proposition \ref{prop:design} is to show what it is \emph{not} doing. Below is an example where the risk level $\za$ would indeed have to be allocated over multiple local hypotheses.

\begin{eg} \label{eg:P_9} Similarly to Example \ref{eg:P_6}, let $\PPPP=(\{0,1\}^{9})^{1000}$, and consider the below universal graph $\Omega$.
    \begin{figure}[h]
    \centering
        \begin{tikzpicture}[scale=1.5]
      \tikzstyle{graphnode}=[circle, draw, minimum size=0.5cm, inner sep=0pt]
      \node[graphnode] (1) at (0,0) {1b};
      \node[graphnode] (2a) at (1,0.5) {2a};
      \node[graphnode] (2b) at (1,-0.5) {2b};
      \node[graphnode] (3a) at (2,0.5) {3a};
      \node[graphnode] (3b) at (2,-0.5) {3b};
      \node[graphnode] (4a) at (3,0.5) {4a};
      \node[graphnode] (4b) at (3,-0.5) {4b};
      \node[graphnode] (5a) at (4,0.5) {5a};
      \node[graphnode] (5b) at (4,-0.5) {5b};
      \node[graphnode] (6a) at (5,0.5) {6a};
      \node[graphnode] (6b) at (5,-0.5) {6b};
      \node[graphnode] (7a) at (6,0.5) {7a};
      \node[graphnode] (7b) at (6,-0.5) {7b};
      \node[graphnode] (8a) at (7,0.5) {8a};
      \node[graphnode] (8b) at (7,-0.5) {8b};
      \node[graphnode] (9a) at (8,0.5) {9a};
      \node[graphnode] (9b) at (8,-0.5) {9b};
      \node[graphnode] (10a) at (9,0.5) {10a};
      \node[graphnode] (10b) at (9,-0.5) {10b};
    
      \draw[->] (1) --  (2a);
      \draw[->] (1) --  (2b) [dashed];
    
      \draw[->] (2a) -- (3a);
      \draw[->] (2b) -- (3a);
      \draw[->] (2b) -- (3b) [dashed];
      
      \draw[->] (3a) -- (4a);
      \draw[->] (3b) -- (4a);
      \draw[->] (3b) -- (4b) [dashed];
      
      \draw[->] (4a) -- (5a);
      \draw[->] (4b) -- (5a);
      \draw[->] (4b) -- (5b) [dashed];
      
      \draw[->] (5a) -- (6a);
      \draw[->] (5b) -- (6a);
      \draw[->] (5b) -- (6b) [dashed];
      
      \draw[->] (6a) -- (7a);
      \draw[->] (6b) -- (7a);
      \draw[->] (6b) -- (7b) [dashed];
      
      \draw[->] (7a) -- (8a);
      \draw[->] (7b) -- (8a);
      \draw[->] (7b) -- (8b) [dashed];
      
      \draw[->] (8a) -- (9a);
      \draw[->] (8b) -- (9a);
      \draw[->] (8b) -- (9b) [dashed];
      
      \draw[->] (9a) -- (10a);
      \draw[->] (9b) -- (10a);
      \draw[->] (9b) -- (10b) [dashed];
      
    \end{tikzpicture}
    \caption{The universal graph $\Omega$ in Example \ref{eg:P_9}}
    \label{fig:P_9}
    \end{figure}
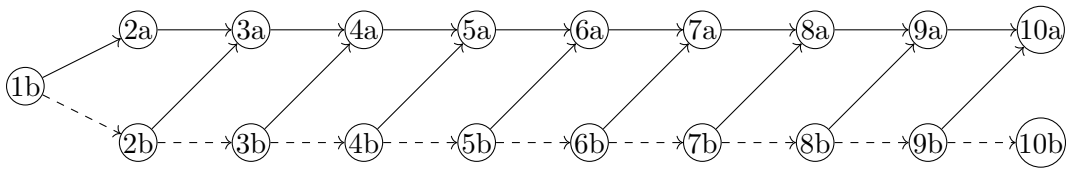\\
    Here there is no set subgraph $G$ or boundary $\partial G$ -- indeed we will see that this is the problem -- but the null hypothesis we wish to reject is $H_\star = $ ``The process ends at $10$b.'' If we define $P_i$ to be the number of $i^{\text{th}}$ component $1$s in the draws from $P\in \PPPP$, then the rule we use to traverse the graph is 
    \[
        \RRR_P((i)\text{b}) =
        \begin{cases}
            (i+1)\text{b} & \ztext{if} P_i\geq 200;\\ 
            (i+1)\text{a} & \ztext{if} P_i<200.
        \end{cases}
    \]
    There is only one path leading from node $1$b to $10$b, which must pass through all of the dashed edges between nodes $(i)$b and $(i+1)$b. If we define $H_i=$``$P_i\geq200$'' for $1\leq i \leq 9$, we may express $H_\star$ as a conjunction of the $H_i$:
    \[
        H_\star = \bigwedge_{1\leq i \leq 9} H_i.
    \]
    If we were to naively take the converse approach to Proposition \ref{prop:design}, we might be tempted to reject $H_\star$ if and only if we can reject a single one of the $H_i$:
    \[
        R_\star = \bigcup_{1\leq i\leq 9}R_i,\ztext{where} \PP(R_i\mid H_i)\leq \za.
    \]
    However we would find that $R_\star$ can have size as large as $9\za$ in this case: because there are $9$ opportunities for the $R_i$ to falsely reject $H_i$, and one false rejection is sufficient to make $R_\star$ falsely reject $H_\star$, the risk level of all the $R_i$ can add up in the worst case.\4 
    We verify this computationally in Table \ref{table:examples_3}. With the state space $\PPPP$ and transition rule $\RRR$ defined above, we take the same experimental approach as in Example \ref{eg:P_6}: randomly sample $100$ of the points in $P$ without replacement, and let $\pi_i$ be the number of $1$s in the $i^{\text{th}}$ components among those points. Then a valid $\za = 0.05$ rejection rule for $H_i$ is
    \[
        R_i = \{P\in\PPPP\colon \pi_i<14\}.
    \]
    Using the cdf of the hypergeometric distribution, we again find the size of this rule is actually $\za\approx 0.039$, in the sense that $\PP(P\in R_i\colon P_i =200)\approx 0.039$. In the worst case, if we synthetically pick $P$ as an element from $\PPPP_0 = \{P\in \PPPP\colon P_i = 200 \; \forall\: i\}$, then the true process should end in vertex \Circled{10b} (i.e. $H_\star$ would be true), but each of the dashed edges along the path to \Circled{10b} gets rejected by the RLA with probability $0.039$. Therefore
    \[
        \PP(R_\star\mid P \in \PPPP_0) \approx 1-(1-0.039)^9\approx 0.3\leq 9\za = 0.351.
    \]
This is verified computationally in the second row of Table \ref{table:examples_3}, under the `Naive Design' entry.
    There are several ways that we could fix this experimental design to test $H_\star$ at level $\za$:
    \begin{enumerate}
        \item We could set lower risk levels $\za_i$ for each of the $H_i$, such that $\sum_{1\leq i \leq 9}\za_i = \za$. \4 
		In the above setting, we might raise the bar to reject $H_i$ as long as $\pi_i \leq 10$, which happens with probability $0.004$ when $P_i =200$. The cost of this is a more conservative test: we end up failing to reject $H_\star$ more often, even when it is false. The results of this test when $P$ is taken from $\PPPP_0$ are also in Table \ref{table:examples_3}, under `Risk-Budgeted Design.'
	\item We could also choose to audit a single one of the $9$ edges on the path from node \Circled{1b} to \Circled{10b} at level $\za$. This amounts to the same idea as solution 1., where we set all but one of the $\za_i$ equal to $0$; if we do not audit an edge, we also do not introduce a risk of making a Type I error. \4 
		For example, if the only edge we choose to audit is the one connecting \Circled{1b} to \Circled{2b}, then we are back in the setting where we can apply Proposition \ref{prop:design}, with the subgraph $G$ made up of nodes $1$b, $2$a, $\ldots$, $10$a, and $\partial G = \{(1\text{b}, 2\text{b})\}$. We show the result of this approach in the last row of table \ref{table:examples_3}.
	\item Alternatively, we could flip the script, and define $H_\star$ as the statement ``The process ends at vertex \Circled{10a}'' instead. In this setting we could apply Proposition \ref{prop:design} with the subgraph $G$ made up of nodes $1$b, $\ldots$, $10$b, and our set of local nulls to reject would contain one assertion for each of the edges that \emph{aren't} dashed in Figure \ref{fig:P_9}. This is statistically equivalent, for this example, to the approach in 2.
    \end{enumerate}
\end{eg}
\begin{table}[h]
	\centering
\begin{tabular}{|l|l|r|r|r|}
\hline
Example & Design Used & Trials & Number of Type I Errors & Empirical Risk Level \\
\hline \hline
\ref{eg:P_6} & Proposition \ref{prop:design} & 50000 & 0 & 0.000000 \\ \hline \hline
\ref{eg:P_9} & Naive Design & 50000 & 15035 & 0.300700 \\ \hline
\ref{eg:P_9} & Risk-Budgeted Design & 50000 & 1757 & 0.035140 \\ \hline
\ref{eg:P_9} & Proposition \ref{prop:design} & 50000 & 1959 & 0.039180 \\ \hline

\end{tabular}
\caption{Computational Verification of the Risk Level for Examples \ref{eg:P_6} and \ref{eg:P_9}} \label{table:examples_3}
\end{table}
Examples \ref{eg:P_6} and \ref{eg:P_9} suggest the following rule of thumb to determine when a rejection rule $R_\star$ for $H_\star$ is designed correctly. If $H_\star$ was true but $R_\star$ rejected it, then the true process corresponding to $P$ must have followed a deterministic path leaving $G$. If we can point to at least one single edge $\ve$ along this path whose local hypothesis $H_\ve$ was definitely rejected at level $\za$ as a part of $R_\star$, then the audit was correctly constructed, and we just made a legitimate Type-I error. Otherwise, if there are multiple such edges along the path, and it is not clear which of their local hypotheses was rejected at level $\za$ as a part of $R_\star$, then we have failed to adequately allocate our risk level $\za$.
\subsection{Variance Bounds for Sparse Samples.}\label{sec:variance_bounds}

The actual parameters we end up sampling for our audits, when we project them onto our local nulls, end up looking a lot like the previous examples; they are functions of the means of (almost) coin flips with a low success rate $p$. When the sample size $n$ and the success rate $p$ are both small, we will see sparse samples where the sample variance may not be an accurate approximation of the true variance. The common rule of thumb for such a case is to require $np(1-p)\gg1$; in our implementations we will stay on the safer side and only use the sample variance when our sample shows more than $20$ non-zero entries (which rarely happens).\4 
When we run into sparse samples, we will fall back on a hypergeometric bound for variance. In the hypergeometric setting, we have a population of $N$ items, $K$ of which record a `success event' -- in our setting, a ballot comparison discrepancy. We sample $n$ of these items without replacement, and we let $X$ be the number of successes in our sample. The hypergeometric distribution tells us 
\begin{equation}
	\PP(X = k) = \frac{{K\choose k}{{N-K}\choose{n-k}}}{{{N}\choose{n}}}.\label{eq:hypergeometric}
\end{equation}
When $K$ is unknown but we have recorded a sample of size $n$ with $k$ successes, we can use this distribution function to build a one-sided confidence interval for $K$; we will follow a procedure called `pivoting the cdf,' stated generally in Theorem 9.2.2 of \textcite{casella_statistical_2002}. \textcite{bartroff_optimal_2022} also explores this method as applied to hypergeometric variables specifically; the exact method we use is outlined at the start of their Section 6. We re-state a version of this Theorem 9.2.2 below, then apply it in our previous context:
\begin{customnumbered2}[(Casella and Berger.)]
	For each $\zt_0\in\Theta$, let $A(\zt_0)$ be the acceptance region of a level $\za$ test of $\H_0\colon \zt=\zt_0$. For each $x\in \mathcal{X}$, define a set $C(x)$ in the parameter space by
	\[
		C(x) = \{\zt_0\colon x\in A(\zt_0)\}.
	\]
	Then the random set $C(x)$ is a $1-\za$ confidence set.
\end{customnumbered2}
In our case, the parameter space $\Theta=[N+1]$ corresponds to the possible values of $\zt= K$, the sample space $\mathcal{X}=[n+1]$ corresponds to our recorded number of successes $X$ in our sample of size $n$, our null hypotheses are $\H_0\colon K=K_0$, our alternative hypotheses are $\H_1\colon K< K_0$, so that large $X$ is evidence to accept $\H_0$, hence we build our acceptance regions as follows: 
\begin{equation}
	A(K_0) = \{k_0\in [n+1]\colon \PP_{K_0}(X\leq k_0)\geq \za\}.\label{eq:A(K)}
\end{equation}
These probabilities are just the cdf of the hypergeometric:
\[
	\PP_{K_0}(X\leq k_0) = \sum_{i=0}^{k_0} \frac{{K_0\choose i}{{N-K_0}\choose{n-i}}}{{{N}\choose{n}}}.
\]
These are stochastically decreasing in $K$ -- when $K_0$ increases and $k_0$ is fixed, it becomes less likely to see $X\leq k_0$ -- hence our acceptance regions are nested:
\[
	A(N)\subseteq A(N-1)\subseteq \ldots \subseteq A(1)\subseteq A(0) = [n+1].
\]
Because of this, our confidence sets $C(x)$ will always be an interval starting at $0$:
\[
	C(x) = \{K_0\colon x\in A(K_0)\} = \{0, \ldots, K_u\}.
\]
Hence $K_u$ is a stochastic upper bound on $K$. In practice, it is the largest $K_0$ such that $x\in A(K_0)$; referring to Equation (\ref{eq:A(K)}), this means by definition that
\begin{equation}
	K_u(x) = \max\{K\in [0,N]\colon \PP_{K}(X\leq x) \geq \za\}.\label{eq:Ku}
\end{equation}
Now we can use this bound to get a straightforward stochastic bound on the variance of $X$:
\begin{align*}
	\Var(X) = \EE(X^2)-\EE(X)^2 \leq \EE(X^2) = \frac{K}{N}, \ztext{hence} \PP(\Var(X)\leq \frac{K_u}{N})\geq 1-\za.
\end{align*}
When the sparse variable $X$ is not a coin flip, we can still use this variance bound as long as $X$ is bounded above by some $B$; in this case the bound is replaced by $(B^2K_u)/N$.\4 
Of course, since this is only a stochastic bound, this means we must allocate some of our global risk level $\za$ to the process of finding a variance bound for our sparse samples. Our local nulls corresponding to each edge $\ve\in\pG$, which previously just depended on some margin $\MMM$, and might have read as ``$\H_\ve\colon L\leq \MMM\leq U$'', have now become composite -- something like
\[
	\H_{\ve}\colon \text{``}\forall\; i\colon \Var(X_i)\leq K_{i}/N \ztext{\textbf{and} in such a case}L\leq\MMM\leq U.\text{''}
\]
where the $X_i$ are the (usually many) hypergeometric variables which are used to build the confidence interval for $\MMM$. For a $3$-seat Meek contest, the worst type of such a local null will use $17$ such `assorter' variables $X_i$! So we will have to be careful about how we budget our risk. In general, we will split our overall risk level as $\za = \za_0 + \za_K$, and use the allocated risk $\za_K$ for the variance bounds we use. There are several way we could allocate the risk $\alpha_K$ to our multiple $X_i$:
\begin{enumerate}
	\item Split $\alpha_K$ into as many parts as there are $X_i$, and bound each $X_i$ with its corresponding risk level;
	\item Make a global variable $Y=\sum_i |X_i|$, treating it as a hypergeometric where ``$Y>0$'' is a success, and find a hypergeometric bound $K$ on the number of successes for $Y$, using all of the allocated risk $\za_K$. Then, use the same bound on variance it induces for each of the $X_i$. That is to say, in the worst case, the number of successes in the population for each individual $X_i$ is bounded above by the number of successes in the population for any of the compilers at all, and the variance bound follows.
	\item We could also do something similar to 2., but with $Y$ corresponding to the event ``there is a ballot-comparison discrepancy (not necessarily detected by the assorters $X_i$).'' This would make for worst bounds on variance, but it has the advantage of being re-useable across every local null we have to reject.
\end{enumerate}
In practice, we end up using option 2. for our implementation. This avoids further breaking $\za_K$ into small pieces, at a cost for the sharpness of the resulting bound. In some cases, the sampled $X_i$ might be $0$, whereas $Y$ can be greater than $20$. It is very plausible that option 1. will be more competitive in the long run, but deciding how to budget the risk $\za_K$ among all of the $X_i$ will be a problem of some complexity. To be optimal, a solution would have to look at the expected scales of the $X_i$ and their impact on the overall margin. A good reference for this approach of budgeting risk as a function of the expected outcome is \textcite{ek_adaptively_2023}.
\subsection{Audit Graphs.}\label{sec:audit_graphs} 
In view of the approach we outline to audit algorithmic election rules in Section \ref{sec:experimental_design}, we had better formalize the graph $\Omega$ and the Meek rule $\RRR_P$ we will use to traverse it. Let the set of candidates $C=[M]$ and the number of seats to fill $m\in \NN$ be fixed.
\begin{defn}
A \textbf{depth-}$\boldsymbol{t}$ \textbf{election state} is an ordered pair of subsets $(H, W)\in \PPP(C)^2$ such that $H\cap W=\varnothing$, $|W|\leq m$, and $|C - H|=t$. We call $H$ the set of \textbf{hopeful} candidates in the state, and we call $W$ the set of \textbf{elected winners}.\4 
The \textbf{degree of an election state} is $\deg(H, W) = |W|$.\4
The \textbf{projection map onto an election state} $S=(H, W)$ is $\pi_S\colon R_C\to R_{H\cup W}$ defined as $\pi_S = \pi_{H \cup W}$.\4
The \textbf{universal audit graph} $\Omega$ is a directed layered graph whose vertex set is the set of all depth-$t$ election states, for $0\leq t\leq M-m$. The root of $\Omega$ is the election state $(C, \varnothing)$. Layer $t$ of $\Omega$ contains all possible election states of depth $t$. We draw an edge from vertex $S=(H,W)$ to $S'=(H', W')$ when:
\begin{itemize}
    \item $H'\subseteq H$,
    \item $W\subseteq W'\subseteq W\cup H$, and
    \item $|(W'-W) \cup (H - H')|=1$. When the single element $c$ in this set is an element of $W'-W$, we call the edge $(S,S')$ a \textbf{winner edge}; note in this case that $c\notin H'$, by the definition of election states. Otherwise, when $c\notin W'$, we must therefore have $c\in H-H'$, and we call the edge $(S, S')$ a \textbf{loser edge}. \4 When it is known whether the edge $\ve =(S, S')$ is a winner or loser edge, we sometimes use the notation $\ve = (S,c)$ to emphasize the candidate $c$ that gets elected/eliminated in this edge.
\end{itemize}
Finally, we define an \textbf{audit graph} as any connected subgraph $G\subseteq \Omega$ which contains the root of $\Omega$, as well as at least one depth $M-m$ election state $S$. We call the graph \textbf{coherent} when any two election states in $G$ coming from that layer have the same winner set.
\end{defn}
In view of the discussion under Algorithm \ref{defn:meek_rule}, we might question the decision to include winner edges in the construction of audit graphs at all. There is one convenient reason to do so, and another more necessary. First, including winner edges makes this construction more general -- we will need fewer modificiations to apply this framework to the WIGM rule, where election rounds are truly necessary.\4 
The second, more serious reason to track winners as part of the election states is that we need to generalize the Meek rule to apply to every vertex in $\Omega$ in such a way that the keep factors $k(i)$ are auditable in every round. In order to calculate variances for the magins between candidates, we will want exact mathematical formulas for the limiting keep factors that the iterative process in step \Circled{2} of Algorithm \ref{defn:meek_rule} converges to. This is possible, but only as long as the set of current winners is known. So, we will define a generalized Meek rule that takes in a prescribed set of winners $W\subseteq C$ as part of its input, and calculates keep factors only for those candidates using the exact formulas.\4 
One awkward consequence of this approach is that occasionally the audit graph will contain election states $S=(H, W)$ where some of the winners in $W$ don't actually make quota in state $S$ according to the CVR. For example, vertex \Circled{3a} of the audit graph in Figure \ref{fig:ward9-meek} of our case study represents an election state where Henry Anderson is in the winner set $W$, despite the CVR recording his final tally as $901.54$ in that round, whereas quota was $904.07$.\4 
The prescription of Meek rule as defined in \ref{defn:meek_rule} in such a case would simply be to truncate the keep factor of Anderson at $1$. But again, this is not a statistically viable option for us, because the variances we obtain by using the exact keep factor formulas only apply at the actual solution of those polynomials. Instead, we opt to use an `extended Meek rule' which uses the exact formulas to calculate keep factors no matter what -- even when the solved keep factors end up above $1$ (which happens in so-called `irregular' election states).\4 
This can look quite irregular indeed! When a winner has a keep factor above $1$, they effectively `steal' votes out of the tallies of candidates down-ballot from their $\fpv$ (`first place vote') ballots. If those $\fpv$ votes exhaust after ranking the winner, they exhaust with a negative weight, thus artificially inflating the number of votes in the election, and quota by extension. In some particularly pathological election states with multiple `fake winners,' the keep factors can blow up to $+\infty$! Although in practice our audit graphs will usually stay quite far away from the parts of $\Omega$ where this can happen. The keep factors we will run across only occasionally go above $1.05$, and rarely above $1.1$.\4
Our mathematical argument in this case is the following: such a state is irregular according to the CVR, and may or may not be so according to the true profile BAL. If it was also irregular according to BAL, then we know for sure that the true Meek process will never visit that state in the audit graph, so whatever edges our auditing process allow to leave from that irregular state just make the audit more conservative; we are giving ourselves more work, not less, by considering this irregular state. If the state was not irregular according to BAL, then the solutions to the exact keep factor equations should be non-singular in a neighborhood of BAL, which our sampled parameters will get us into, so that our estimates of margin variances should hold.\4 
We should spare a word as well for how we construct the audit graphs $G$ we end up using for our audits; even if this process is of no importance for the actual statistical framework we apply to it (we might as well treat $G$ as provided by a third party), it might still be of interest as a general-purpose construction.\4 
What we do is that we fix a `Least Auditable Margin' $\text{LAM}\in\NN$, which we think of as the smallest margin we feel confident we would be able to audit with the sample size we are planning. Then, starting at the root of $\Omega$, we calculate the tallies in each election state, and we allow every outcome which could be possible if margins of fewer than LAM votes were allowed to flip. This is similar to giving an adversary a budget of $\text{LAM}/2$ votes they are allowed to flip (whichever way they want) and telling them to try their worst; any outcome they might be able to achieve will be included in our graph $G$. Crucially, we reset the adversary's budget in every new election state we reach. They might flip one set of $\text{LAM}/2$ votes to get from the root to a state in the first layer of $G$, at which point we would reset the CVR and they would be allowed to pick a new (possibly different) set of $\text{LAM}/2$ votes to flip.
\subsection{Instant Keep Factors.}\label{sec:keep} 
To write down the formulas for the instant keep factors, we will need to track some quantities of interest in each projected profile $P_H = \pi_H\circ P$.
\begin{defn} (Profile Projections.) \label{defn:profile_projections}\4
	Given a profile $P$ on candidates $H\subseteq C$, a distinguished $L$-ranking $s\in R_C$, and a distinguished candidate $c\in H$:
	\begin{itemize}
		\item We define the \textbf{first-}$\boldsymbol{L}$ restriction map $\first_L\colon R_H\to R_H$ as the map that truncates each ranking to its first $L$ positions:
			\[
				\first_L(r) = \begin{cases}
					r\Big|_{[L]} & \ztext{if} \len(r) > L; \\
					r &\ztext{if} \len(r) \leq L.
				\end{cases}
			\]
		\item We define the \textbf{initially-like-$\boldsymbol{s}$} tally $T_s(P)$ as the number of rankings $r$ `inside of' $P$ whose first-$L$ restriction is $s$:
			\[
				T_s(P) = |(\first_L \circ P)^{-1}[s]|.
			\]
			N.b.: if $s$ ranks a candidate not in $H$, this is automatically $0$.
		\item We define the \textbf{exactly-like-}$\boldsymbol{s}$ tally $t_s(P)$ as the number of times a ranking $r$ in $P$ is $s$:
			\[
				t_s(P) = |P^{-1}[s]|.
			\]
	\item We distinguish $t_\varnothing(P)$ by giving it the notation $g(P)$ (or just $g$, when $P$ is clear) and calling it the \textbf{ghost number} of $P$: this is the number of rankings in $P$ that indicate no preference in $H$.
	\item We use the following shorthand to represent rankings $r$ in the subscript of these tallies: when $\len(r)=L$, and the positions in $r$ are $r(0)= r_0, \ldots, r(L-1)=r_{L-1}$, we use the notation
		\[
			T_{r_0,\ldots, r_{L-1}} \ztext{and} t_{r_0,\ldots, r_{L-1}} \ztext{to represent} T_r \ztext{and} t_r.
		\]
	For example, $T_{5,3,12}$ represents the number times a vote $r= P(i)$ has $r(0)=5$, $r(1)=3$, and $r(2) = 12$. We will occasionally drop the commas if all numbers involved are small.
\item Finally it will be helpful to distinguish the \textbf{ending-in-$\boldsymbol{c}$ rankings} $R_{H;c}\subseteq R_H$ as 
	\[
		R_{H,c} = \{r\in R_H : r(\len(r)-1)= c\}.
	\]
	N.b. that $R_{H; c}$ contains the singleton ranking $r = \{(0,c)\}$, among others; a vote that lists only $c$ is still a vote ending in $c$.
\end{itemize}	
\end{defn}

Given a profile $P\colon [N]\to R_C$ on candidates $C$ and an election state $(H, W)$, we set $P_S = \pi_S(P)$, and define the number of \textbf{active votes} $N'$ in the election state is:
\begin{equation}
	N' = N - \sum_{r\in R_W}\left(\prod_{i=0}^{\len(r)-1}\Big(1-k(r(i))\Big)\right)t_{r}(P_S),\label{eq:active_votes}
\end{equation}
where the \textbf{keep factors} $k\colon W\to (0,+\infty]$ are the least positive solutions to the system of equations:
\begin{equation}
	\forall \; w\in W : \frac{N'}{m+1}+\zve = k(w)\sum_{r\in R_{W; w}}\left(\prod_{i=0}^{\len(r)-2}\Big(1-k\big(r(i)\big)\Big)\right)T_r(P_S).\label{eq:instant_keep}
\end{equation}
This system of equations might have no such solution in general; we will have to worry about this when designing our extended Meek rule. This is also not the first time someone solves for the convergent value of keep factors directly -- Meek himself did this for degree $\leq 4$ election states in Section 7 of his original paper \parencite{meek_nouvelle_1969}. Presently we confirm that when the solution exists, it agrees with the solution found by Algorithm \ref{defn:meek_rule}:
\begin{customnumbered}
	For a fixed profile $P\in\PPPP_{H}$, let $k\colon H\to(0,1]$ be the convergent keep factors obtained by applying step \Circled{2} of Algorithm \ref{defn:meek_rule} with a tolerance $\omega$, and define $W\defeq \{c\in H: k(c)<1\}$. If there exists a solution $k'$ to Equations (\ref{eq:active_votes}) and (\ref{eq:instant_keep}) applied to equation state $S = (H\backslash W, W)$, then this solution is unique, and it agrees with $k$ up to a factor of $\omega$, in the sense that for all $c\in W$,
	\[
		|k(c)-k'(c)| < \frac{\omega}{q}<\omega.
	\]
\end{customnumbered}
\textbf{Proof.}\4 
It suffices to note that such a solution $k$ is by design a fixed point of the Meek algorithm; in such a case Theorem 1 of \textcite{hill_algorithm_1987} already shows it is unique, and Theorem 2 of that same paper proves it is `within tolerance' of the iterative solution from the algorithm. \qed \4
With the instant keep factors defined, the `extended Meek rule' we described in Section \ref{sec:audit_graphs} is just to use the instant formulas (when possible) rather than the iterative process described in step \Circled{2} of Algorithm \ref{defn:meek_rule} to calibrate the keep factors. Given that the rest of the algorithm -- including how the keep factors are used to elect/eliminate candidates -- remains the same, this extended Meek rule is guaranteed to produce the same winner set as the original rule, for any fixed profile.\4 
However, when we apply this instant keep factor formula to election states that would not be naturally be visited by the Meek rule, it is possible to see keep factors greater than $1$ -- we call such states `irregular,' in the following sense:
\begin{defn}
	We call an election state $S=(H,W)$ \textbf{regular} if the system in (\ref{eq:instant_keep}) has a solution $k\colon W\to[0,1]$ where all the keep factors are between $0$ and $1$. We call it \textbf{degenerate} if there is no finite positive solution $k\colon W\to[0,+\infty)$.
\end{defn}
There is reason to fear that, when we solve the keep factors for election states not visited by the true Meek rule in our audit graph, the instant keep factor solutions for that state might vary discontinuously with respect to the profile projections $T_r$ and $t_r$ used to solve them. In such a case, the confidence intervals we build for the margins between candidates in this election state, which depend on this variation of the keep factors with respect to these projections, would not be valid. On the one hand, we can offer at least a modicum of comfort here: this can not happen as long as the election state is regular.
\begin{customnumbered2}\label{thm:comfort}
	The (unique) solution $k\colon W\to[0,1]$ for a regular election state is continuous with respect to the profile projections $T_r$ and $t_r$ used to construct it.
\end{customnumbered2}
\textbf{Proof.}\4 
Fix a degree-$j$ election state $S=(H,W)$, call $C=H\cup W$, and consider the profiles $P\in\PPPP$ on candidates $C$. Let $d=|R_C|$ and define $\TTT = \RR^d$, so that the Equations in (\ref{eq:instant_keep}) can be thought of as describing a system of polynomial equations $\vec{F}$:
\begin{equation}
	\vec{F}\left(\vec{k},\vec{t}\right) = \vec{0}\ztext{for} \vec{t}\in \TTT \ztext{and} \vec{k}\in \RR^j.\label{eq:vector_form}
\end{equation}
Finally let $\TTT_{\mathsf{reg}} \subseteq \TTT$ correspond to the set of profile projections that make $S$ regular; by definition, for every $\vec{t}\in\TTT_{\mathsf{reg}}$, there exists a unique solution $\vec{k}\in [0,1]^j$ such that (\ref{eq:vector_form}) holds. This means we can define a function $\vec{k}\colon \TTT_{\mathsf{reg}}\to [0,1]^j$ which maps every regular $\vec{t}$ to the unique $\vec{k}$ that satisfies (\ref{eq:vector_form}). We will show this function is continuous.\4
To see this, let $\vec{t}_n\to \vec{t}$ in $\TTT_{\mathsf{reg}}$, and define the sequence $\vec{k}_n\defeq \vec{k}\left(\vec{t}_n\right)$. This sequence lives in a compact set $[0,1]^j$, which implies it has a convergent subsequence $\vec{k}_{n_i}\to \vec{k}_\infty$. By continuity of $F$, we see
\[
	0 = \vec{F}\left(\vec{k}_{n_i}, \vec{t}_{n_i}\right)\to \vec{F}\left(\vec{k}_\infty,\vec{t}\right).
\]
By definition of $\vec{k}$, this implies that $\vec{k}_\infty = \vec{k}(t)$. Using the same reasoning, any other convergent subsequence of the $\vec{k}_n$ has the same limit. Hence the function $\vec{k}$ is continuous.\qed \4 
On the other hand, we have some confirmation that these fears are well-founded. We occasionally need to include irregular states in our audit graphs (for example, vertex \Circled{3a} in the audit graph for our case study). We have no guarantee from Theorem \ref{thm:comfort} that the keep factor solutions for these election states are well behaved; and indeed we find that in general it is possible to find states that are arbitrarily close to being regular where these solutions are not well-behaved.
\begin{customnumbered2}\label{thm:discomfort}
	For arbitrary $m\geq 3$ and arbitrarily large $N$, there exists a set of candidates $C$ containing a subset $W=\{w_0, w_1\}$ and a profile $P\colon [N]\to R_C$ such that the election state $S= (W, C\backslash W)$ is regular, but switching two votes in $P$ would make $S$ degenerate.
\end{customnumbered2}
\textbf{Proof.}\4 
We construct a profile $P$ on $m$ candidates. Label the candidates as $w_0=0, w_1=1, 2, \ldots, m-1$. Define 
\[
	x = \left\lceil \frac{N}{m+1}+\zve \right\rceil \ztext{and} y = \left\lfloor\frac{N}{m+1}-\zve\right\rfloor.
\]
Define the rankings $r_0 = \{(0,0), (1,1)\}$ and $r_1=\{(0,1),(1,0)\}$ -- these are rankings that list one winner as their first preference, the other as their second, and no other candidates subsequently. For $2\leq j\leq m-1$, define $r_j = \{(0,j)\}$.\4 
Construct the profile $P$ by assigning $r_0$ to the first $x$ rankings, $r_1$ to the next $x$ rankings, then $y$ rankings to each $r_j$ for $2\leq j< m-1$, and finally assigning the remaining $N-2x-(m-3)y$ rankings to $m-1$. Clearly $S$ is a regular election state for $P$, because $0$ and $1$ have a quota's worth of votes to begin with by construction.\4 
However, if we `take away' one $r_0$ ranking and one $r_1$ ranking (replacing them, for instance, with rankings $r_{m-1}$), so that there are now $x' = x-1$ rankings $r_0$ and $r_1$ in the new profile $P'$, we find the election state is degenerate. By symmetry, the keep factors of candidates $0$ and $1$, if they exist, must be equal; so label them as $k$. Given $P'$, the instant keep factors in (\ref{eq:instant_keep}) would be solutions of the polynomial $Ak^2+Bk+C$, where 
\begin{align*}
	A&= mx' & B&=-2mx' & C&=N+(m+1)\zve -x'.
\end{align*}
The discriminant of this polynomial is $\Delta = 4mx' (mx'- (N+(m+1)\zve -x'))$. This is non-negative if and only if $(m+1)x'>N+(m+1)\zve$, i.e. if and only if $x' > N/(m+1)+\zve = q$. But by construction, $x'$ is less than quota, so $S$ is degenerate for $P'$.\qed\4
In view of this result, the best we can do is offer anecdotal evidence why we might not expect to see such turbulent election states in real-world profiles. Firstly, this phenomenon can only occur for election states of degree 2 or higher -- this pathology can not arise when there are $<2$ winners. \4
Secondly, the pathology relies on having many voters rank both `insecure winners' one after the other; if a sufficient number of voters rank another hopeful candidate, or a secure winner, or no candidate at all after an insecure winner, the election state stops being degenerate. In Australia, for example, Above-the-Line voting means that when voters rank multiple winners in a row, the first (several) winners in that ranking are very secure, and only the last winner may be insecure.\4 
In particular, these two points imply that we are safe if we are auditing an election with $m=3$ seats, as long as one of the two winners in every degree 2 vertex of the audit graph is secure; this is certainly the case for all of the graphs we used in our Sections \ref{sec:portland} and \ref{sec:australia}.\4 
Finally, we outline some modifications of the Meek rule in Section \ref{sec:future} that might resolve these pathologies, while still being naturally auditable via our graph-based method.
\section{Local Null Audits.}
Rather than discuss the statistical methods used to audit our local nulls in full generality, we will use our case study to show those details in practice. Once these details are laid out, we will finally be able to give estimates of the \textbf{Average Sampling Numbers (ASN)} of our methods in section \ref{sec:ASN}.
\subsection{General Setup.}\label{sec:general_setup}
As outlined in section \ref{sec:experimental_design}, our strategy is to reject one null hypothesis for each edge in a distinguished set $\pG\subseteq \vE_\Omega$. Here we define these local null hypotheses, and outline how we project our sample of ballot-CVR pairs onto the rejection rules for each of these local nulls.\4 
In each of the following subsections, we will re-formulate the local nulls and their rejection rules explicitly in the context where the degree of the election state $S$ our local null concerns is small, but we can also formulate these rejection rules generally. To do this, we borrow the notion of an \textbf{assorter} from the RLA literature -- cf. \textcite{stark_sets_2020}, Section 2.

\begin{defn} (Assorters.)\4 
	When $\BAL, \CVR \in \PPPP$ are fixed $N$-voter profiles on candidates $H\subseteq C$, and $r\in R_H$ is a distinguished $L$-ranking, we define two \textbf{discrepancy assorter} maps
	\[
		\zd_r, \zl_r\colon[N]\to \{-1,0,+1\},
	\]
	as follows:
	\begin{align*}
		\zd_r(i)&=\begin{cases}
       	 -1 & \text{if} \; \first_L\circ \CVR(i) = r \ztext{and} \first_L\circ \BAL(i) \neq r;\\
        1 & \text{if} \; \first_L\circ \CVR(i) \neq r \ztext{and} \first_L\circ \BAL(i) = r;\\
        0 & \text{otherwise.}
    \end{cases} \\ 
    \zl_r(i)&=\begin{cases}
       	 -1 & \text{if} \; \CVR(i) = r \ztext{and} \BAL(i) \neq r;\\
        1 & \text{if} \; \CVR(i) \neq r \ztext{and} \BAL(i) = r;\\
        0 & \text{otherwise.}
    \end{cases}
	\end{align*}	
\end{defn}

These assorters measure how far off the profile projections $T_r(\CVR)$ and $t_r(\CVR)$ are from their true values $T_r(\BAL)$ and $t_r(\BAL)$ (cf. \ref{defn:profile_projections} for definitions), in the following sense:
\[
	T_r(\BAL) = T_r(\CVR) + \sum_{i=0}^{N-1}\zd(i) \ztext{and} t_r(\BAL)= t_r(\CVR)+ \sum_{i=0}^{N-1}\zl(i)
\]
Since the projections of the true profile $\BAL$ determine the path taken by the extended Meek rule through the universal graph $\Omega$, our strategy is to use the projections of the $\CVR$ as an `educated guess' for their values, and use our random sample to correct those guesses by estimating the quantities 
\[
	\mu_r = \overline{\zd_r} = \ds\frac{1}{N}\sum_{i=0}^{N-1} \zd(i)\ztext{and}\nu_r =\ds \overline{\zl_r} = \frac{1}{N}\sum_{i=0}^{N-1} \zl(i).
\]
To distinguish these `educated guesses' from the projections of the true profile, we distinguish them with a \v{} accent:
\begin{align*}
	\vT_r &\defeq T_r(\CVR) \ztext{and}  \vt_r \defeq t_r(\CVR), \ztext{whereas}\\ 
	T_r &\defeq T_r(\BAL) \ztext{and}  t_r  \defeq t_r(\BAL).
\end{align*}
The quantities we end up auditing for our local nulls $\H_\ve$ of an election state $(H,W)$ are margins $\MMM_{i,j}$ between the tallies $\TTT_i$ and $\TTT_j$ of candidate pairs $i, j\in H$, as well as margins $\MMM_{i,q}$ between a candidate $i\in H$ and quota $q$. These quantities, originally defined in step \Circled{2} of the Algorithm \ref{defn:meek_rule}, can now be defined more concretely in terms of the profile projections and the instant keep factor solution $k\colon W\to[0,1]$ from Equations (\ref{eq:active_votes}) and (\ref{eq:instant_keep}):
\begin{align*}
	\TTT_i&\defeq \sum_{r\in R_{H\cup W; i}} k(i) \left(\prod_{j=0}^{\len(r)-2}\Big(1-k\big(r(j)\big)\Big)\right) T_r;\\
	\MMM_{i,j} &\defeq \TTT(i)-\TTT(j);\\
	q &\defeq \frac{ N - \sum_{r\in R_W}\left(\prod_{i=0}^{\len(r)-1}\Big(1-k(r(i))\Big)\right)t_{r}}{m+1}+\zve;\\ 
	\MMM_{i,q}&\defeq \TTT_i - q.
\end{align*}
With these quantities defined, we can define the local null hypotheses $\H_\ve$ corresponding to the edges $\ve\in\pG$:
\begin{itemize}
    \item If $\ve=\big(S,c\big)$ is a loser edge, the local null is ``Candidate $c$ loses to every other hopeful $\ell\in H$ and no new candidate gets quota in this round.'' The way we reject it is by either 1) checking $\PP(\MMM_{c\ell}<0)<\za$, where $\ell$ is the recorded loser in $H$, or 2) checking there is some other candidate $h\in H$ that reaches quota in this round, in the sense that $\PP(\MMM_{hq}<0)< \za$. Which of these options we check is something we decide based on the CVR, but we only ever check one of these hypotheses per loser edge. 
\item Similarly, the null for a winner edge $\ve = \big(S, c\big)$ is ``Candidate $c$ makes quota in $S$, and has the highest tally among hopefuls in that round.'' We reject it by either 1) checking that $\PP(\MMM_{cq}>0)\leq \za$, or $2)$ checking there is some other hopeful $h\in H$ such that $\PP(\MMM_{ch}>0)<\za$.
\end{itemize}
\begin{minipage}{0.49\textwidth}
    For the rest of this section, we outline the specifics of these rejection rules for degree-0, 1, and 2 election states, and demonstrate how we apply them to succesfully audit the Perth-Kinross profile from our case study in Section \ref{sec:case_study} re-imagined as a Meek election. As we do this, it will be helpful to refer back to its audit graph, which we re-draw in Figure \ref{fig:perth_audit_tree}. \4 
    The original profile for this election contained $N = 3689$ votes, to which we add $150$ `ghost' votes to emulate votes that were invalidated before the election started, but might be added back in if they are sampled and found valid after all.
\end{minipage}\hfill
\begin{minipage}{0.49\textwidth}
    \centering\begin{tikzpicture}[scale=1]
      \tikzset{
        graphnode/.style={circle, draw, minimum size=0.5cm, inner sep=0pt},
        edgelabel/.style={font=\scriptsize, align=center}
      }

      \node[graphnode] (n1) at (0,4) {1};
      \node[graphnode] (n2) at (0,3) {2};
      \node[graphnode] (n3) at (-1,2) {3a};
      \node[graphnode] (n4) at (1,2) {3b};
      \node[graphnode] (n5) at (0,1) {4};
      \node[graphnode] (n6) at (0,0) {5};
      \node[graphnode] (n7) at (0,-1) {6};

      \draw[->]  (n1) -- node[edgelabel, right]{Elect\\Alan Livingstone} (n2);
      \draw[->]  (n2) -- node[edgelabel, left]{Elect\\Henry Anderson\\} (n3);
      \draw[->]  (n2) -- node[edgelabel, right]{Eliminate\\George Hayton\\} (n4);
      \draw[->]  (n3) -- node[edgelabel, left]{Eliminate\\George Hayton} (n5);
      \draw[->]  (n4) -- node[edgelabel, right]{Elect\\Henry Anderson} (n5);
      \draw[->]  (n5) -- node[edgelabel, right]{Eliminate\\Andrew Dundas} (n6);
      \draw[->]  (n6) -- node[edgelabel, right]{Elect\\Wilma Lumsden} (n7);
    \end{tikzpicture}
        
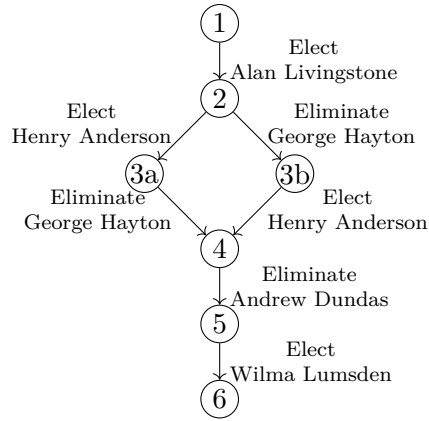
\captionof{figure}{\centering The audit tree for an \\
    auditable margin of 40 votes.}
    \label{fig:perth_audit_tree}
\end{minipage}\4
We pick a sample size of $n=767$ votes, or about 20\% of the total number of ballot papers; this is on the larger side for an \textbf{Average Sampling Number (ASN)}, to account for the tight margin of $47$ votes we must audit in the last round of the election. To conduct this synthetic audit, we artifically noise the CVR by perturbing a fraction $\eta = 0.05$ of the rankings, including the artifical ghost ballots, in one of 4 ways, following a standard procedure established in the Appendix of \textcite{blom_raire_2019}: either
\begin{enumerate}
	\item Replace a random candidate in the ranking with another candidate not listed in the ranking;
	\item Insert a random candidate that does not appear in the ranking into a random position in the ranking;
	\item Randomly swap the positions of two candidates in the ranking; or 
	\item Remove a random candidate from the ranking.
\end{enumerate}
Finally we will use a risk level of $\za = 0.05$ for this synthetic audit.

\subsection{Degree-0 Audits.}

When the winner set of an election state $S=(H, \varnothing)$ is empty, there are no keep factors to calibrate, and margins are straightforward to compute by hand. We need only worry about reducing quota after exhausting votes that listed only candidates in $H\compc$. This new quota simplifies as 
\[
    q  = \frac{N-g}{m+1}+\zve, 
\]
Here there are only two discrepancy assorters of interest $\zd_c$ and $\zl_g\eqdef\zl_\varnothing$. We re-iterate their formal definitions in this context:
\begin{align*}
	\zd_c(i)&=\begin{cases}
        -1 & \text{if} \; \fpv_S(\CVR(i)) = c \ztext{and} \fpv_S(\BAL(i))\neq c;\\
        1 & \text{if} \; \fpv_S(\CVR(i)) \neq c \ztext{and} \fpv_S(\BAL(i)) = c;\\
        0 & \text{otherwise.}
    \end{cases}\\
		\zl_g(i)&=\begin{cases}
        -1 & \text{if} \;\pi_S\circ\CVR(i) = \varnothing \ztext{and} \pi_S\circ\BAL(i)\neq \varnothing;\\
        1 & \text{if} \;\pi_S\circ\CVR(i) \neq \varnothing \ztext{and} \pi_S\circ\BAL(i)= \varnothing;\\
        0 & \text{otherwise.}
    \end{cases}
	\end{align*}
Using these assorters we can correct our guesses from the CVR and write a formula for the true exhaust total $g$ and true tallies $\TTT_c$ for $c\in H$:
\begin{align*}
    \TTT_c = T_c &= \vT_c + \sum_{i=0}^{N-1} \zd_c(i) = \vT_c + N\mu_c, \ztext{where} \mu_c = \overline{\zd_c} = \frac{1}{N}\sum_{i=0}^{N-1} \zd_c(i);\\
    g &= \vg + \sum_{i=0}^{N-1} \zl_g(i)= \vg + N\mu_g, \ztext{where} \nu_g = \overline{\zl_g} = \frac{1}{N}\sum_{i=0}^{N-1} \zl_g(i).
\end{align*}
If we have estimators $\hmu_c$ and $\hnu_g$ for $\mu_c$ and $\nu_g$, we can use them to construct estimators of our margins of interest: for fixed $c, \ell\in H$, we have
\begin{align*}
    \hMMM_{c\ell} &= (\vT_c-\vT_\ell) + N(\hmu_c-\hmu_\ell);\\
    \hMMM_{cq} &= \left(\vT_c - \frac{N-\vg}{m+1} - \zve\right) + N\left(\hmu_c + \frac{\hnu_g}{m+1}\right).
\end{align*}
In the degree-0 case, we could combine the two estimators $\hmu_c, \hmu_\ell$ into a single estimator $\hmu_0= \hmu_c-\hmu_\ell$. This would make the SHANGRLA compilers of \textcite{stark_sets_2020} applicable to this margin. Likewise for $\hM_{cq}$. We opt not to do this, to keep all of our auditing framework consistent about its usage of the delta method.\4 
Instead, for $\MMM_{cq}$, we pick $\zvt=(\mu_c,\nu_g)$ and take the gradient of $\MMM_{cq}$:
\[
	\nabla_{\zvt}\MMM_{cq}=[\partial_{\mu_c}\MMM_{cq},\partial_{\nu_g}\MMM_{cq}]\tranT = \left[N, \frac{N}{m+1}\right]\tranT
\]
Then, if $\Sigma_{\zvt}$ is the covariance matrix corresponding to $\zvt$ (with the hypergeometric variance bounds on the diagonal as discussed in Section \ref{sec:variance_bounds}), the delta method says that asymptotically,
\begin{equation}
	\Var(\hMMM_{cq})\approx \nabla_{\zvt}\MMM_{cq}\tranT \Sigma_{\zvt} \nabla_{\zvt}\MMM_{cq}\Big|_{\zvt = (\hmu_c, \hnu_g)}.\label{eq:delta_method}
\end{equation}
Finally, we multiply by the Simple Random Sampling Without Replacement (SRSWOR) correction factor $\frac{N-n}{N-1}$ and take a square root to get the standard error of the margin, which we use to build a one-sided confidence interval.\4 
The margin for $\MMM_{c\ell}$ is similar, only with $\zvt = (\mu_c, \mu_\ell)$.

\begin{eg} When we audit vertex \Circled{1} in our case study, one the main local nulls we have to reject is ``Livingstone does not have quota at the start of the election.'' If we call Livingstone candidate $c$, then our assorters record the following:
	\begin{itemize}
	    \item $\zd_c$ contains 5 entries with a $-1$, and 4 entries with a $+1$; the other 758 entries are all $0$.
	    \item $\zl_g$ contains 2 entries with a $-1$, and 4 entries with a $+1$; the other 761 entries are all $0$.
	    \item One of the non-zero entries in $\zl_g$ overlaps with a non-zero entry of $\zd_c$, such that in total there are $15$ ballot dicrepancies relevant to this margin; the induced variance bound is $K/N = 0.0328$, which we divide by $n$ to bound the variance of the sample mean.
	    \item The resulting delta method matrix multiplication from \ref{eq:delta_method} yields
		    \begin{align*}
			    \begin{bmatrix}
				    3839 \\
 					 959.75 \\
				\end{bmatrix}\tranT
				\begin{bmatrix}
  0.000043 & -0.000002 \\
  -0.000002 & 0.000043 \\
\end{bmatrix}
				\begin{bmatrix}
				    3839 \\
 					 959.75 \\
				\end{bmatrix}\approx 657.56
		    \end{align*}
	    \item Finally we multiply by the SRSWOR factor $(N-n)/(N-1)\approx .8$ and square root to get a $\text{SE}\approx22.94$. After multiplying by the appropriate $z$-score and adding the sampled margin, we get a lower bound of $149.51$ on $\MMM_{cq}$. This bound being positive confirms that Livingstone does make quota with more than $95\%$ confidence.
    \end{itemize}
	The only other nulls we would have to check in vertex \Circled{1} is that Livingstone beats every other candidate in that round (i.e. that no other candidate would get elected before him).
\end{eg}
\subsection{Degree-1 Audits.}
In the setting where there is a single elected candidate $w$ in the election state, candidate-to-candidate margins have the form
\begin{align}
    &\hMMM_{c\ell} = (\vT_c-\vT_\ell)+N(\hmu_c-\hmu_\ell) +\bigg(1-\underbrace{\frac{\vC_u-N\hmu_u}{\vC_v+N\hmu_v}}_{\hk}\bigg)(\vT_{wc}-\vT_{w\ell}+N(\hmu_{wc}-\hmu_{w\ell}))\nonumber\\
    &= (\vT_c-\vT_\ell+\vT_{wc}-\vT_{w\ell})+N(\hmu_c-\hmu_\ell+\hmu_{wc}-\hmu_{w\ell})-\frac{\vC_u-N\hmu_u}{\vC_v+N\hmu_v}(\vT_{wc}-\vT_{w\ell}+N(\hmu_{wc}-\hmu_{w\ell})).
\end{align}
where
\begin{align*}
	\vC_u & = N-\vg -\vt_w + (m+1)\zve,  & \hmu_u & = \hnu_g+\hnu_w,\\
	\vC_v & = (m+1)\vT_w-\vt_w, & \hmu_v & = (m+1)\hmu_w - \hnu_w. 
   \end{align*}
The candidate-to-quota margin can be similarly simplified:
\[
	\hMMM_{cq} = (\vT_c + \vT_{wc})+N(\hmu_c-\hmu_{wc}) - \frac{\vC_u-\hmu_u}{\vC_v + \hmu_v}(\vT_{wc}+\vT_w + N(\hmu_{wc}+\hmu_w))
\]
We use the same delta method to give one-sided confidence intervals for these margins. For $\MMM_{c\ell}$ the parameters are $\zvt = (\mu_c, \mu_\ell, \mu_{wc}, \mu_{w\ell}, \nu_g, \nu_w, \mu_w)$. For $\MMM_{cq}$ the parameters are $\zvt = (\mu_{c}, \mu_{wc}, \nu_g, \nu_w, \mu_w)$. The non-linear dependence of the margins on these parameters highlights the need for the delta method.
\begin{eg}
	In election state \Circled{2}, one of the loser edges $\ve\in\pG$ is the edge corresponding to Dundas, who we refer to as $c$ here. The recorded loser in this election state is Hayton, whom we refer to as $\ell$ here. The null hypothesis corresponding to this loser edge is ``$\MMM_{c\ell} < 0$.'' \4 
The original CVR margin between these candidates in this election state is $\check{\MMM}_{c\ell}=94.1464$. After correcting this margin with our sampled parameters, it decreases to an estimate of $\hMMM_{c\ell}=81.2074$.\4
There are $24$ discrepant ballots among those sampled, yielding an individual variance bound of $K/N<0.000064$. The delta method calculation in equation (\ref{eq:delta_method}) is too large to print, but it yields a variance bound of $\Var(\hMMM_{c\ell})<1959.4421$. After multiplying by our SRSWOR factor we get a standard error for our margin of $\hSE \approx39.6026$, yielding a lower bound of $\hMMM_{c\ell}>16.0668>0$ for our 95\% confidence interval, allowing us to reject this local null.\4 
The other nulls we would have to check at this vertex are that no candidate other than Anderson makes quota in this election state, and that no hopeful candidate loses to Hayton head to head.
\end{eg}

\subsection{Degree-2 Audits.}

Without loss of generality, label the two winners in a degree-$2$ election state as $1,2$. In this setting the instant keep factor equations in (\ref{eq:instant_keep}) can be solved to isolate $k_1=k(1)$ and $k_2=k(2)$. We find that the keep factor $k_2$ is the root of the polynomial $P_2(\zvt, k) = A_2(\zvt)k^2 + B_2(\zvt)k+C_2(\zvt)$, where
\begin{eqnarray}
A_2 & = & 
T_2t_{12} - T_{12}t_2 + T_2t_{21}+T_{21}t_{12}+T_{21}t_2+T_{21}t_{21}-(m+1)\Big(T_{12}T_{21}+T_2T_{21}\Big)\label{eq:A1} \\
B_2 & 	= & -T_1t_{12}-T_1t_2+T_{12}t_2-T_2t_1-T_2t_{12}-T_1t_{21}-T_2t_{21}+(N-g)(T_{21}-T_{12}) - T_{21}t_1\nonumber \\
    & &\phantom{(m+1)} -2(T_{21}t_{12}+T_{21}t_2+T_{21}t_{21})\label{eq:B1}\\
    & & \phantom{(m+1)}+(m+1)\Big(T_1T_{12}+T_1T_2+T_{12}T_{21}+T_2T_{21}-T_{12}\zve+T_{21}\zve\Big)\nonumber \\
C_2 & = & -(T_1+T_{21})\Big(N-g-t_1-t_{12}-t_2-t_{21}+(m+1)\zve\Big) \label{eq:C1}
\end{eqnarray}
The keep factor $k_1$ is similarly a root of $P_1(\zvt, k) = A_1(\zvt)k^2 + B_1(\zvt)k+C_1(\zvt)$, whose coefficients just have the indices $1$ and $2$ switched. We solve these equations numerically with the sampled parameter $\zvt$ to find $\hk_1$ and $\hk_2$.\4 
For the candidate-to-candidate margin $\MMM_{c\ell}$, our parameters are
\begin{equation}
	\zvt =  \big[\nu_g,\mu_{1},\mu_{2},\mu_{12},\mu_{21},\nu_{1},\nu_{2},\nu_{12},\nu_{21},\mu_c, \mu_\ell, \mu_{1c}, \mu_{1\ell}, \mu_{2c},\mu_{2\ell},\mu_{12c}+\mu_{21c}, \mu_{12\ell}+\mu_{21\ell}\big]\label{eq:degree_2}
\end{equation}
N.b. that the sum $\zd_{12c}+\zd_{21c}$ is still bounded above by $1$, since it is not possible for these assorters to simultaneously be $+1$ or $-1$. Likewise for $\zd_{12\ell}+\zd_{21\ell}$.\4
In finding $\nabla_{\zvt}\MMM_{c\ell}$, we will need to take some partials $\partial_{\theta_i}k$. We use the Implicit Function Theorem to do this:
\[
    \frac{\partial k}{\partial\theta_i}=-\frac{\partial_{\theta_i} P(k,\zvt)}{\partial_k P(k,\zvt)}
    = -\left.\frac{(\partial_{\theta_i}A)k^2 + (\partial_{\theta_i}B)k+(\partial_{\theta_i}C)}{2Ak+B}\right|_{\zvt, k = \zht, \hk}
\]
Finally the candidate-to-candidate margin has the following form:
\begin{align*}
	\hMMM_{c\ell} & = (\vT_c - \vT_\ell)+ N(\hmu_c-\hmu_\ell)\\
		      &+(1-\hk_1) \left( \vT_{1c}-\vT_{1\ell}+N(\hmu_{1c}-\hmu_{1\ell})\right) \nonumber\\
				   &+(1-\hk_2) \left( \vT_{2c}-\vT_{2\ell}+N(\hmu_{2c}-\hmu_{2\ell})\right) \nonumber 	\nonumber\\
				   &+(1-\hk_1)(1-\hk_2) \left( \vT_{12c}+\vT_{21c}-\vT_{12\ell}-\vT_{21\ell}+N(\hmu_{12c}+\hmu_{21c}-\hmu_{12\ell}-\hmu_{21\ell})\right) \nonumber
\end{align*}
We could similarly write down a symbolic form for $\hMMM_{cq}$, but degree-2 candidate-to-quota margins are not necessary when $m=3$; it is sufficient to check that the last recorded winner of the election beats every remaining hopeful candidate in every degree-2 vertex.
\begin{eg}
	Vertex \Circled{3A} is the only irregular election state in our graph; according to the CVR, the keep factors would be $k_1 = 0.8130$ and $k_2=1.0029$ for the recorded winners Livingstone and Anderson. After substituting our sampled parameters into the coefficient equations (\ref{eq:A1}) - (\ref{eq:C1}), these become regular again; we get $\hk_1 = 0.814$ and $\hk_2 = 0.9887$.\4
	The tightest margin we have to audit in this election state is again to check that Dundas does not lose to Hayton; the recorded margin between them (using the irregular keep factors) is $94.0624$; this margin decreases to $81.6326$ after applying our sampled parameters, and it is bounded below by $14.4745>0$ according to our delta method, allowing us to reject this local null. Each of the Implicit Function Theorem derivatives $\partial_{\theta_i}k$ are bounded above by $10^{-1}$ in our example, and the number of relevant ballot comparison discrepancies is $26$, leading to a variance bound of $K/N<0.0520$.
\end{eg}
\subsection{Average Sampling Numbers.}\label{sec:ASN}
\begin{figure}[h]
	\centering
		\includegraphics[width = \textwidth]{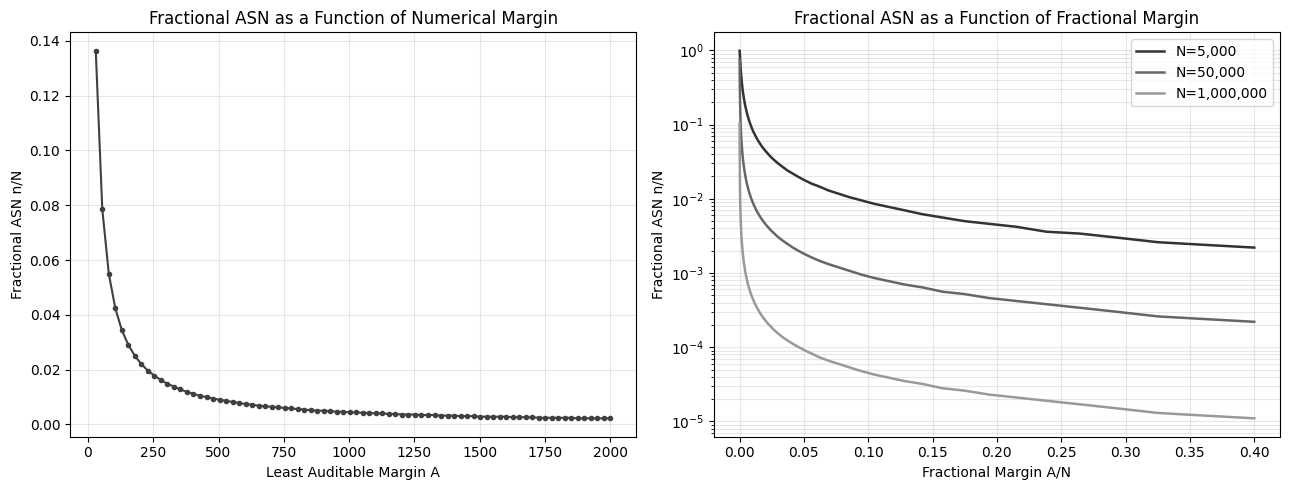}
		\caption{ASN as a function of Least Auditable Margin $A$ when $\za=.05$.}
		\label{fig:ASN}
\end{figure}

With the mechanics of local null audits established, we are equipped to discuss the expected minimal sample size $n$ needed to successfully audit a given election, which we refer to as its \textbf{Average Sampling Number (ASN)}. This number depends mainly on the edge $\ve\in\pG$ in the chosen audit graph whose rejection rule has the smallest margin. In this section we will treat the ASN as a function of this \textbf{Least Auditable Margin (LAM)} $\zm$, and we will only consider the case where the noise level $\eta=0$. This reduction will help give us a sense of the scale of the ASN as a function of the smallest margin in the election.\4
In the case where $\eta=0$, we find that the variance bound $K/N$ from Section \ref{sec:variance_bounds} is asymptotically proportional to $\frac{1}{n}$ in the regime where $n\ll N$. This in turn implies that the fractional ASN $n/N$ needed for a succesful RLA will depend only on $\zm$, not on its scale relative to $N$. That is to say, the cost as a fraction of the profile size to audit a margin of $\zm=100$ votes is the same regardless of whether $N=5000$ or $N=1,000,000$; or in other words, a margin of $1\%$ costs comparatively less to audit in a larger profile. These results are underscored in Figure \ref{fig:ASN}.\4 
This seems like good news: it is reasonable to expect larger profiles will have numerically larger LAMs, in which case a successful RLA will have to sample a smaller fraction $n/N$ of the profile. 
We spend the rest of this section mathematically justifying Figure \ref{fig:ASN}.
\begin{customnumbered} \label{prop:ASN}
	In the setting where $\eta=0$ and we budget the risk-limit as $\za=\za_0+\za_K$ to audit a numerical Least Auditable Margin $\zm$ of the $N$ votes in a given profile, and there are $d$ assorter parameters $\zvt$ used to build the margin in question, the asymptotic relationship between the Average Sampling Number $n$ and $A$ is 
	\[
		\frac{n}{N} \approx \frac{z_{1-\alpha_0}\sqrt{d\log(1/\za_K)}}{\zm}
	\]
\end{customnumbered}
\textbf{Proof.}\4 
When we sample $n$ out of the population of $N$ ballots without replacements to find $X=0$ ballot discrepancies, and $K$ is unknown a priori, the hypergeometric probability in Equation (\ref{eq:hypergeometric}) simplifies to
\begin{align*}
	\PP_K(X= 0) & = \frac{{{K}\choose{0}}{{N-K}\choose{n}}}{{{N}\choose{n}}} = \frac{\frac{(N-K)!}{n!(N-K-n)!}}{\frac{N!}{n!(N-n)!}}\\ 
		    & = \frac{(N-K)(N-K-1)\cdots(N-K-n+1)}{N(N-1)\cdots(N-n+1)}\\
		    & = \prod_{j=0}^{n-1}\left(1-\frac{K}{N-j}\right)
\end{align*}
Taking logarithms and using the approximation $\ln(1+u)= u + O(u^2)\approx u$ in the regime where $u\ll 1$, i.e. $K\ll N-n$, we get
\begin{align}
	\ln \PP_K(X=0) & = -\sum_{j=0}^{n-1} \frac{K}{N-j}\nonumber\\ 
		       & = -K\sum_{j=0}^{n-1}\left(\frac{j}{N(N-j)}+\frac{1}{N}\right)\nonumber\\ 
	\ln \PP_K(X=0) &=-K\left(\frac{n}{N}+\sum_{j=0}^{n-1}\frac{j}{N(N-j)}\right) \label{eq:Pk}
\end{align}
Now we note the sum is $O(n^2/N^2)$ in the regime $n\ll N$, since $N-j\geq N-n$, hence
\[
	0\leq \sum_{j=0}^{n-1}\frac{j}{N(N-j)}\leq \frac{1}{N(N-n)}\sum_{j=0}^{n-1}j = \frac{n(n-1)}{N(N-n)}= O\left(\frac{n^2}{N^2}\right).\tag{as $n\ll N$}
\]
Recalling from Equation (\ref{eq:Ku}) that $K_u$ is the maximum value of $K$ such that $\PP_K(X=0)\geq \za_K$, we set the left hand side of Equation (\ref{eq:Pk}) equal to $\ln(\za_k)$ and we substitute $K=K_u$ into the right hand side, to find
\[
	\ln(\za_K) = -K_u\left(\frac{n}{N}+O\left(\frac{n^2}{N^2}\right)\right) \approx -K_u\frac{n}{N}. \tag{as $n\ll N$}
\]
In other words, our ballot-level stochastic variance bound is $K_u/N\approx{\ln(1/\za_K)/n}$. Dividing by $n$ again to turn it into the variance of the sample, and ignoring the finite population correction factor in the regime $n\ll N$, we find the width of a one-sided $(1-\za_0)$-confidence interval resulting from the sum of $d$ of these variance bounds is 
\[
	w =N\cdot  z_{1-\za_0}\sqrt{\frac{d\ln(1/\za_K)}{n^2}}.
\]
The result follows from setting $w=\zm$ and solving for $n$.\qed
\section{Experimental Results.}\label{sec:results}
We can run this auditing framework on several databases of Cast Vote Records (CVRs) for past elections that used Ranked Choice Voting to measure success rate and viability\footnote{
	Replication code available \href{https://github.com/EdouardHeitzmann/meek.git}{\underline{here}}.
}. We use the same methodology as we did in our case study to produce these results: starting from the CVRs, we introduce a noise level $\eta$ by perturbing a corresponding fraction of the ballots, synthetically emulating a paper ballot record, which the CVR is a noisy approximation of. We noise these ballots using the same procedure described at the end of Section \ref{sec:general_setup}.\4 
None of these elections were originally run using the Meek rule -- all of those included here used the WIGM rule, with either $m=2,3,4,$ or rarely $m=5$ winners. We re-imagine all of them as $2$- or $3$-seat Meek elections, using the same CVRs. We also introduce some `ghost ballots' to each profile -- empty rankings which list none of the initial candidates -- to simulate ballots that were spoilt before the election started. These ballots have no impact on the election itself (in Meek, they do not even change the quota), but there is a possibility our noise turns one of these ghosts into a valid ballot. We introduce one such ballot for every $100$ ballots in the profile.\4 
In this setting where the noise level $\eta$ between our two profiles is non-zero, we will consider the \textbf{Average Sampling Number (ASN)} to be the smallest sample size of the profile that will lead to a successful audit over $90\%$ of the time.\4 
In view of our discussion in Section \ref{sec:ASN}, we expect ASNs to decrease as profile sizes increase. To confirm this, we will run our ASNs on three `scales' of elections: $N\approx 5,000$, $N\approx 50,000$, and $N\approx 1,000,000$.
\subsection{Small Elections: Scotland.}\label{sec:scotland}
\begin{figure}[H]
	\centering
		\includegraphics[width = .85\textwidth]{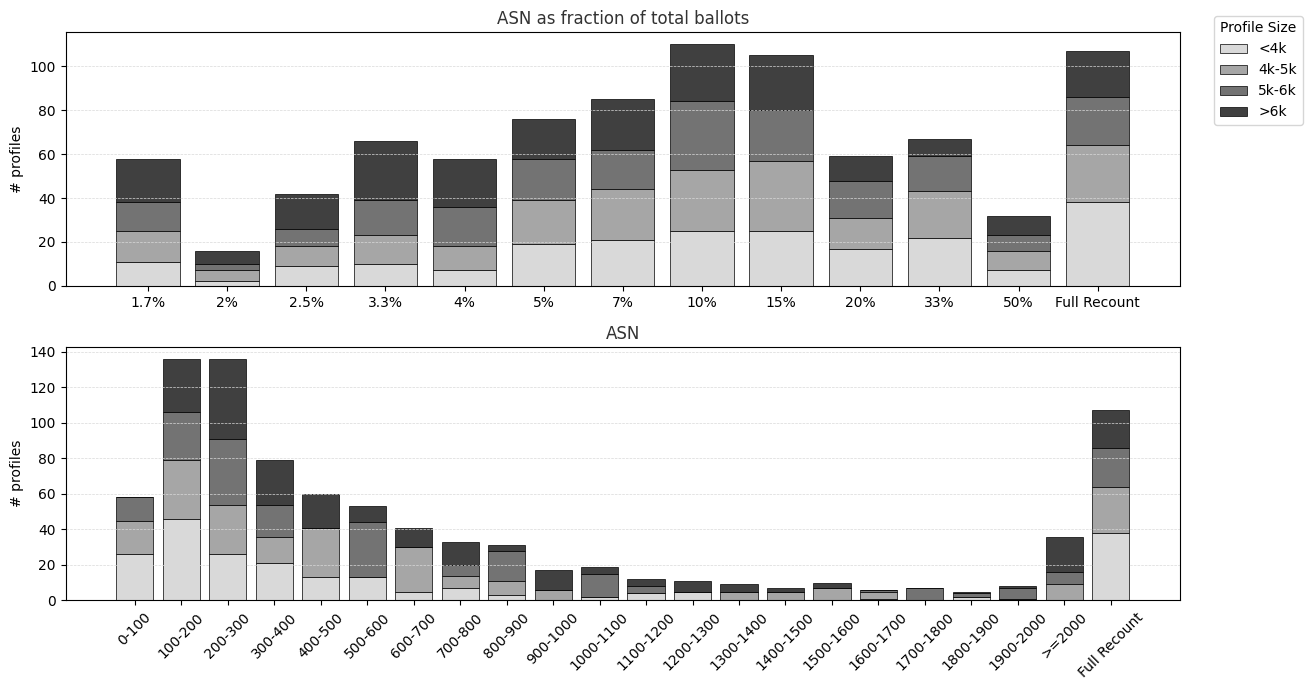}
		\caption{Average Sampling Number across $881$ Scottish STV elections, $\za = .05$}
		\label{fig:alpha5}
\end{figure}

\begin{figure}[h]
	\centering
		\includegraphics[width = .7\textwidth]{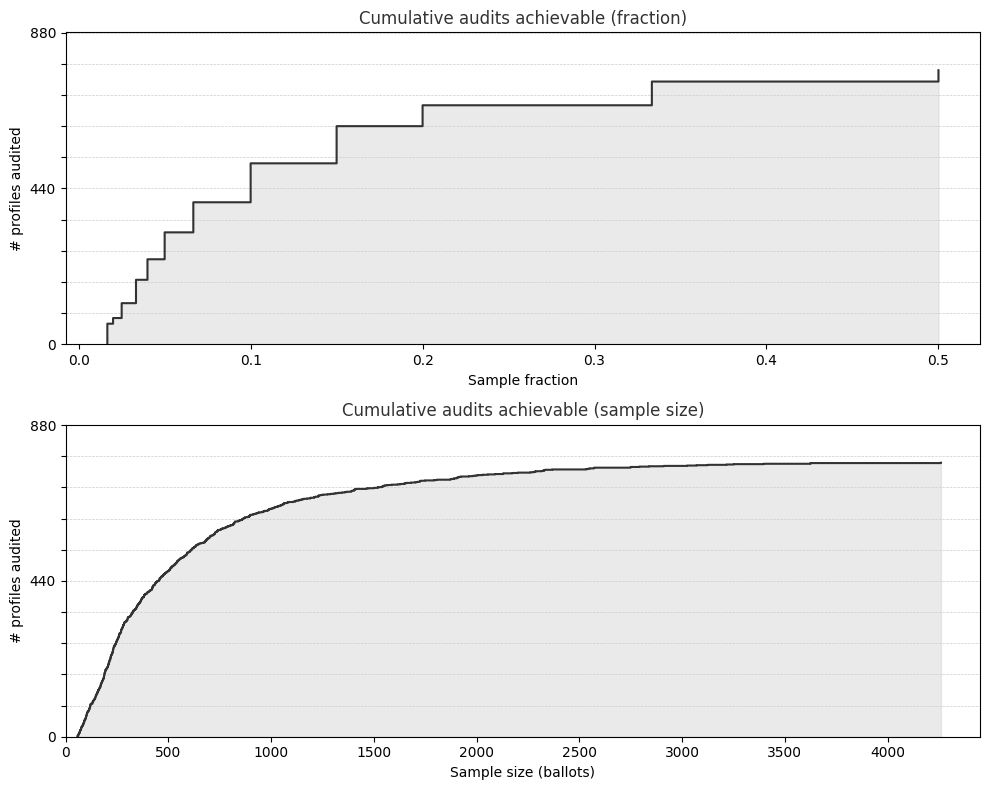}
		\caption{\centering Number of Scottish elections auditable\\ with an ASN at or below given threshold, $\za=.05$.}
		\label{fig:a5_cumulative}
\end{figure}

\begin{figure}[H]
	\centering
		\includegraphics[width = .7\textwidth]{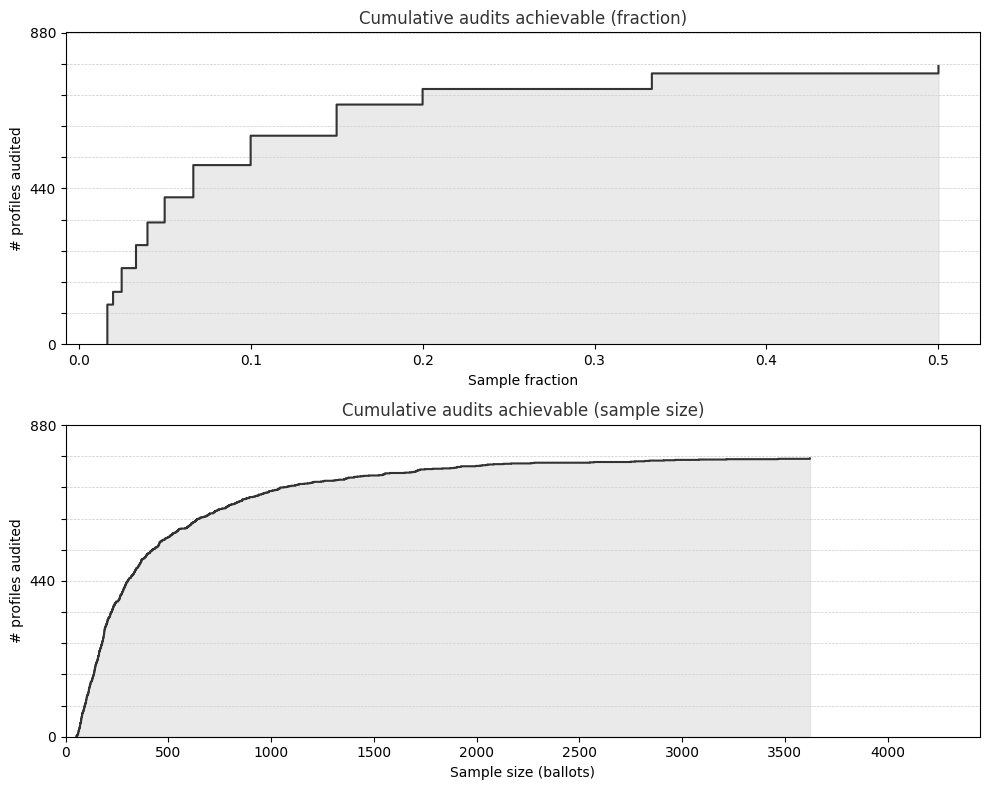}
		\caption{\centering Number of Scottish elections auditable\\ with an ASN at or below given threshold, $\za=.1$.}
		\label{fig:a10_cumulative}
\end{figure}

Scotland has been using WIGM STV for its city councils since 2007. This has produced a rich database of 1068 ranked-choice voting profiles for small-to-medium sized elections (one of which was the profile for our Perth-Kinross case study). \4
We ran our auditing methods on all profiles in this database that had $|C|= 6,7,8,9,$ or $10$ candidates running. This includes $881$ profiles, of which $1$ was a race for $m=2$ seats, $439$ profiles were races for $m=3$ seats, $439$ more profiles were races for $m=4$ seats, and $1$ profile was a race for $5$ seats. The electorate for these elections is comparatively small: the average number of voters across all audited profiles is around $5,248$.\4 
For these audits, we used a noise level of $\eta =2\%$; this is still a lot higher than recorded real-world noise levels, which are typically a fraction of a percent for political elections. We start by auditing these profiles with a risk level of $\za = .05$, of which we allocate $\za_K = .005$ to the construction of the variance bounds discussed in Section \ref{sec:variance_bounds}. We plot the resulting ASNs in Figure \ref{fig:alpha5}, and the number of profiles auditable as a function of sample size in Figure \ref{fig:a5_cumulative}.\4 
\mbox{}\\
To run batch audits on this large dataset and find ASNs, we used a 2-step greedy search process. First, we build audit graphs with iteratively larger minimal auditable margins $M$, continuing as long as the degree-3 vertices of the graph all have the same winner set as that recorded by the CVR. Second, we find the smallest sample size $n$ that will lead to a successful RLA for the given audit graph over $90\%$ of the time. It's worth noting that this greedy process is often not optimal; the search process only checked a handful of auditable margins and sample sizes, both of which could be pushed further by an individual tailoring them to each specific profile.\4
Here is a verbal summary of these results: $76.6\%$ of the profiles, or $675/881$, were auditable with an ASN of $\leq 30\%$ of the number of votes in the profile. Of the $107$ profiles that were not auditable with any ASN of at most $50\%$ of the profile, $62.1\%$ of them recorded a margin of less than $0.5\%$ of the profile between the last winner of the election and their strongest opponent. In American city council elections, it is typical for such a margin to trigger an automatic full recount.\4
We repeated these audits with an increased risk level of $\za=.1$, of which we allocated $\za_K=.02$ to the variance bounds from Section \ref{sec:variance_bounds}. The cumulative results are shown in Figure \ref{fig:a10_cumulative}: only $12$ more profiles became auditable, but the ASNs needed to successfully audit each profile decreased slightly.
\lstset{language=Python, basicstyle=\ttfamily\small, frame=single,
  keywordstyle=\color{RoyalBlue}\bfseries,
  commentstyle=\color{PineGreen}\ttfamily, showstringspaces=false}
  \lstset{numbers=left, numberstyle=\tiny, stepnumber=2, numbersep=5pt}
\subsection{Medium Elections: Portland.}\label{sec:portland}
The city of Portland switched its city council election to use WIGM STV for the first time in 2024. This election was split into $4$ districts, each of which elected $3$ councilors, for a total of $12$. These districts had a turnout of $42,678,\, 77,024,\, 84,356,$ and $76,566$ voters respectively\footnote{
N.b. that we artificially add `ghost votes' to each of these profiles proportional to 1\% of their weight, so the actual turnouts we use for the audits are inflated by a factor of 1.01x.%
}, for an average district size of $70,156$. For our purposes, we re-imagine them as Meek elections, and artifically noise the CVRs with a noise level $\eta = 2\%$, using a risk level of $\za=5\%$ for our audits. Gracefully, the outcome of the election in each of these districts is the same whether we use WIGM or Meek (which was not always the case for the Scotland database, as evidenced by our case study).\4 
The most fragile of these elections was District 1, for which the largest LAM that led to a coherent audit graph was $A=532$, or $1.2\%$ of the profile. We display this audit graph in Appendix \ref{sec:portland_d1} -- it is instructive to look at, because its structure is typical of audit graphs: there is an initially turbulent cloud of low-stakes eliminations (the so-called `elimination cloud'), after which the graph converges back to a single vertex in election state $\Circled{8A}$, at which point the higher-stakes eliminations and elections of the more viable candidates begin. The size of the elimination cloud tends to blow up combinatorially with the number of candidates in the election, and as we will see later this is a recurring computational obstacle for our graph-based RLAs in larger elections. \4 
We find that this audit graph for Portland District 1 is consistently auditable with an ASN of $n=1,077$ ballot comparisons, or around $2.5\%$ of the profile.\4 
The only other Portland profile we contrive to audit close to optimally is District 4, which had a LAM of around $A=2,000$ votes, or $2.6\%$ of the profile. District 4 had 30 candidates (compared to 16 in District 1), and consequently the audit graph corresponding to this LAM is unrecommendably large -- it contains 8,455,380 vertices, more than 95\% of which occur in the first 16 of 31 layers of the graph. For this audit graph the smallest ASN was of $n=386$ votes, or $0.5\%$ of the profile.\4 
Districts 2 and 3 are a lot more stable than the others; the former has a LAM of around $3,500$ votes, or $4.5\%$ of the profile, and the latter has a LAM of $9,000$ votes, or $11.6\%$. The audit graphs corresponding to these margins are too large to compute, leaving us with two options; either use a more modest LAM to get a (less efficient) successful RLA, or pre-emptively remove some of the less viable candidates from each profile, thereby skipping over the elimination cloud stage of the audit graph. The latter option seems like it should be justifiable -- the candidates we remove from each profile all have less than $5\%$ of quota at the time of their elimination in the canonical Meek path through the graph -- but it is not currently mathematically justified. Even so-called `non-viable candidates' -- candidates with fewer mentions than (initial) quota -- might impact the election if their combined presence in a round denies enough votes to a winner to see them eliminated. Mathematically justifying this pre-emptive removal of non-viable candidates would considerably simplify and optimize the framework of graph-based RLAs. We show the ASNs corresponding to both approaches in Table \ref{table:portland}.
\begin{table}[ht]
\centering
\begin{tabular}{|c|c|c|c|c|}
\hline
 & \multicolumn{2}{c|}{\textbf{Rigorous Approach}} & \multicolumn{2}{c|}{\textbf{Optimal Approach}} \\
\hline
Profile & LAM & ASN & LAM & ASN \\
\hline
District 1 & 532 (1.23\%) & 1077 (2.5\%) & 532 (1.23\%) & 1077 (2.5\%) \\
\hline
District 2 & 2,000 (2.4\%) & 311 (0.4\%) & 3,500 (4.1\%) & 253 (0.33\%) \\
\hline
District 3 & 400 (0.47\%)  & 2839 (3.3\%) & 9,000 (11.6\%) & 102 (0.13\%) \\
\hline
District 4 & 2,000 (2.6\%) & 421 (0.55\%)  & 2,000 (2.6\%) & 421 (0.55\%) \\
\hline
\end{tabular}
\caption{LAMs and ASNs for the audits of Portland Districts 1-4.}\label{table:portland}
\end{table}\mbox{}\4
It is worth dwelling for a moment longer on the computational challenge posed by these `elimination clouds.' The graphs we used to audit districts 2 and 3 with a LAM of 2,000 and 400 respectively were made up of 8,244 and 14,955 vertices. Of these vertices, 8,113 (98.4\%) and 14,822 (99.1\%) respectively occurred in the first half of the graph, corresponding to the elimination cloud we describe. In District 4, layer 11 was the largest, containing 1,373,001 vertices, or 16\% of the total, but by layer 23 the graph had collapsed back to a single-vertex layer, indicating the end of the elimination cloud. These clouds bear little impact on the outcome of the election, but currently represent the majority of the computational work of the audit. \4 
Worse, the size of the cloud is dependent mostly on the number of candidates in the election: District 3 had a much more robust election than District 2, but its computationally viable LAM was much lower because the former had 30 candidates while the latter had 22. This emphasizes the need for a mathematically rigorous justification to jump past this elimination cloud; else the best recommendation we will be able to make to keep STV elections secure will be to limit the number of candidates ahead of the election (which is arguably contrary to the spirit of STV elections).

\subsection{Large Elections: Australia.}\label{sec:australia}
Australia has been using STV for its Senate since 1949; they were using digitally tabulated WIGM STV by 1998, and they have published full CVRs for their 2016\footnote{The 2016 election is unusual because it was the result of a `double dissolution' of Parliament, meaning each of Australia's six federal states were electing 12 new senators, rather than the usual 6. This explains why the number of candidates running in each of these states was unusually high that year.}, 2019, 2022, and 2025.\4
Since 1985, the Australian Senate has been made up of 76 seats, 72 of which come from 6 federal states electing 12 senators each for staggered 6-year terms, and 4 of which come from 2 territories electing 2 senators each for 3-year term. Each state- and territory-level senatorial election uses (a version of) WIGM STV. This is the largest-scale use of STV for political elections in the world: the largest of these states, New South Wales, had a turnout of 5,219,341 voters in 2025.\4 
Australia is also singular for its use of ``Above The Line (ATL)'' voting, which allows voters to rank parties instead of candidates if they prefer. When a voter casts an ATL vote, the vote is counted as though it had ranked all candidates of each party listed by the vote, in an order pre-determined ahead of the election. It is common to see $>90\%$ of the electorate cast ATL votes, and consequently Australian STV elections tend to see candidates reach quota much earlier and with larger surpluses than elsewhere.\4 
The two territories that elect 2 senators each every election cycle -- the Northern Territory (NT) and Australian Capital Territory (ACT) -- are small enough (and more importantly, see few enough candidates running for office) that our graph-based approach can usually do a full successful RLA for them. To make the audits computationally viable, we sometimes have to use a smaller LAM than would otherwise be possible; in such cases the resulting ASNs are slightly smaller, but in practice the difference is small as long as the LAM is on the order of thousands. \4
We audit these elections as $m=2$ seat Meek elections, with our usual noise level of $\eta=2\%$ and risk level of $\za = 0.05$. We show the results, as well as the number $M=|C|$ of candidates, the number $N$ of voters (including the 1\% of ghost ballots added to the election) for each election in table \ref{table:territories}.
\begin{table}[h]
\centering
\begin{tabular}{|c|c|c|c|c|c|c|}
\hline
Territory & Year & |C| & N & Optimal LAM & Viable LAM & ASN \\ \hline\hline

\multirow{4}{*}{\makecell{Australian\\Capital\\Territory}}
  & 2016 & 22 & 256,906 & 38,000 (14.7\%) & 2,000 (0.8\%) & 1,027 (0.4\%) \\ \cline{2-7}
  & 2019 & 17 & 272,933 & 35,000 (12.8\%) & 2,000 (0.7\%) & 1,364 (0.5\%)\\ \cline{2-7}
  & 2022 & 23 & 288,069 & 17,000 (5.9\%) & 500 (0.2\%) & $\infty$ \\ \cline{2-7}
  & 2025 & 14 & 296,408 & 40,000 (13.5\%) & 40,000 (13.5\%) & 296 (0.1\%) \\ \hline\hline 

\multirow{4}{*}{\makecell{Northern\\Territory}}
  & 2016 & 19 & 102,516 & 18,000 (17.6\%) & 4,000 (4\%) & 146 (0.14\%) \\ \cline{2-7} 
  & 2019 & 18 & 106,077 & 19,000 (17.9\%) & 3,700 (3.5\%) & 151 (0.14\%) \\ \cline{2-7}
  & 2022 & 17 & 104,653 & 14,000 (13.4\%) & 4,400 (4.2\%) & 139 (0.13\%) \\ \cline{2-7} 
  & 2025 & 17 & 107,875 & 19,000 (17.6\%) & 4,800 (4.45\%) & 134 (0.12\%) \\ \hline 

\end{tabular}

\caption{RLA Results for Australia's Smaller Territories.}\label{table:territories}
\end{table}\mbox{}\\
The only one of these territory elections we weren't able to audit rigorously was the 2022 ACT election. If we use our informal shortcut to remove the 14 least viable candidates from the election before the audit starts (these are all candidates that had less than 4,000 votes as of the time of their elimination in the original election order), the election becomes auditable with an ASN of $n= 146$ votes, or $0.05\%$ of the profile.\4
Australia's six federal states tend to have much larger turnouts and more candidates running for office (often more than 50 candidates). We also don't currently have the technology to audit them as 6-seat elections (although that step forward seems accessible). Rather, we audit them as though they had been 3-seat elections, and we use the same method as we did in Portland to `skip ahead' of the elimination cloud part of the graph -- again, without strict mathematical justification, although the candidates we pre-emptively eliminate tend to have fewer mentions in the profile than the quota they would need to win a seat.\4 
We show the results of our auditing methods applied to Australia's federal states in Table \ref{table:states}. These results include the LAM we used for the audit, the ASN we got out of it, as well as the original number of candidates $|C|$ and the number of remaining candidates $|C'|$ after we `jump over' the elimination cloud.\4 
The results for these Australian audits seem particularly encouraging: for these elections that often have more than a million voters, sampling a few hundred ballots often suffices to perform a successful audit.
\begin{table}[H]
\centering
\begin{tabular}{|c|c|c|c|c|c|c|c|}
\hline
State & Year & |C| & |C'| & N & LAM & ASN \\ \hline\hline

\multirow{4}{*}{\makecell{New\\South\\Wales}}
  & 2016 & 151 & 16 & 4,274,747 & 172,000 (4\%)& 427 (0.01\%)  \\ \cline{2-7}
  & 2019 & 105 & 13 & 4,668,386 & 240,000 (5.14\%) & 194 (0.004\%) \\ \cline{2-7}
  & 2022 & 75 & 11 & 4,822,863 & 113,000 (2.34\%) & 401 (0.008\%)  \\ \cline{2-7}
  & 2025 & 56 & 12 & 5,014,481 & 61,000 (1.22\%) & 1,337 (0.03\%)  \\ \hline\hline

\multirow{4}{*}{\makecell{Victoria}}
  & 2016 & 116 & 14 & 3,477,543 & 13,000 (0.37\%) & 11,591 (0.33\%)  \\ \cline{2-7}
  & 2019 & 82 & 12 & 3,744,918 & 44,000 (1.17\%) & 936 (0.02\%)  \\ \cline{2-7}
  & 2022 & 79 & 12 & 3,834,047 & 211,000 (5.5\%) & 159 (0.004\%)  \\ \cline{2-7}
  & 2025 & 65 & 13 & 4,129,837 & 33,000 (0.7\%) & 2,359 (0.06\%)  \\ \hline\hline

\multirow{4}{*}{\makecell{Queensland}}
  & 2016 & 122 & 17 & 2,731,865 & 95,000 (3.48\%) & 270 (0.01\%)  \\ \cline{2-7}
  & 2019 & 83 & 12 & 2,921,107 & 144,000 (4.93\%) & 208 (0.007\%)  \\ \cline{2-7}
  & 2022 & 79 & 11 & 3,036,380 & 26,000 (0.85\%) & 2,450 (0.08\%)  \\ \cline{2-7}
  & 2025 & 56 & 12 & 3,252,922 & 46,000 (1.41\%)& 896 (0.03\%)  \\ \hline\hline

\multirow{4}{*}{\makecell{Western\\Australia}}
  & 2016 & 79 & 13 & 1,371,310 & 37,000 (2.7\%) & 390 (0.03\%)  \\ \cline{2-7}
  & 2019 & 67 & 13 & 1,459,142 & 85,000 (5.82\%) & 181 (0.01\%)  \\ \cline{2-7}
  & 2022 & 58 & 12 & 1,539,357 & 7,000 (0.45\%) & 5,118 (0.33\%)  \\ \cline{2-7}
  & 2025 & 49 & 12 & 1,636,041 & 11,000 (0.67\%) & 3,265 (0.2\%) \\ \hline\hline

\multirow{4}{*}{\makecell{South\\Australia}}
  & 2016 & 64 & 9 & 1,064,808 & 200,000 (18.78\%) & 177 (0.01\%)  \\ \cline{2-7}
  & 2019 & 42 & 12 & 1,104,303 & 3,000 (0.27\%) & 11,022 (1\%)  \\ \cline{2-7}
  & 2022 & 51 & 12 & 1,137,357 & 47,000 (4.13\%) & 206 (0.02\%)  \\ \cline{2-7}
  & 2025 & 40 & 12 & 1,174,613 & 36,000 (3.06\%) & 335 (0.03\%)  \\ \hline\hline

\multirow{4}{*}{\makecell{Tasmania}}
  & 2016 & 58 & 14 & 341,487 & 7,000 (2.04\%) & 485 (0.14\%)  \\ \cline{2-7}
  & 2019 & 44 & 12 & 355,217 & 12,000 (3.38\%) & 315 (0.09\%)  \\ \cline{2-7}
  & 2022 & 39 & 14 & 364,547 & 227 (0.06\%) & 91,136 (25\%)  \\ \cline{2-7}
  & 2025 & 33 & 12 & 375,463 & 6,000 (1.6\%) & 625 (0.17\%)  \\ \hline

\end{tabular}
\caption{RLA Results for Australia's Federal States.}\label{table:states}
\end{table}

\section{Future Work.}\label{sec:future}
The graph-based approach to algorithmic election RLAs is new, and there is a lot of fertile ground left to explore -- mathematically, algorithmically, and computationally speaking.\4 
One immediate next step is to extend the local null constructions to higher degree election states. In the case of Meek elections, this will require a generalized framework to take partials $\partial_{\theta_i}k$ of higher degree keep factors. The current Implicit Function Theorem framework could remain viable up to $m=5$, but in general it might be a better idea to implicitly differentiate the equations in (\ref{eq:instant_keep}) and solve for the partials numerically. The turbulence of these higher degree solutions might make it more common for the audit graphs to include non-regular election states, which might be further perturbed into becoming degenerate by the auditing process, reducing viable LAMs for these elections. This is already a problem we observed in Portland District 3 -- the high optimal LAM sometimes led to election states where the sampled parameters could not be solved for finite, positive keep factors. In such a case the only viable option is to reduce the LAM until all vertices become close enough to regular.\4 
Another direction to explore is using this graph-based approach on WIGM elections. This will require re-thinking the definition of audit graphs: because the chronology of elections matters in WIGM, the winner set of an election state will have to carry information about the hopeful set at the time of each winner's election. This will make it less common for different branches of the graph to converge back to the same vertices, and as a result the audit graphs will be bushier. In every election state where a winner gets elected, a new branch of the audit graph will have to start, which will never re-join the other branches. \4
This dependence on exact chronology might make WIGM elections more unstable in general -- this is what we observed in the case study from Section \ref{sec:case_study} -- but the relative computational simplicity of WIGM might counteract this, so that it is not clear which of these two election rules will be more auditable in the long run. Especially in cases when the timing of multiple winners is very certain, it might be the case that WIGM is more auditable than Meek. This would be the case, for instance, in Districts 1, 2, and 3 of Portland, where the first winner got elected relatively late and after receiving a large transfer, meaning their quota was quite secure. In District 4, however, Olivia Clark had an initial quota with a small surplus, making the exact timing of her election difficult to pin down in the audit graph -- in the full audit graph we computed, there are 4,097,797 vertices corresponding to her election, each of which would have to create a separate branch if we were using WIGM STV.\4 
Another algorithmic challenge would be to find a statistically rigorous process allowing us to eliminate multiple candidates at the same time to `jump over' the elimination clouds we ran into in our larger audit graphs from Sections \ref{sec:portland} and \ref{sec:australia}. Again, this is a place where Meek might be more structurally suited to the task: when it is possible for a winner to get elected in the middle of the elimination cloud (as was the case in Portland District 4), it seems like there can be no such shortcut in WIGM.\4 %
One outline of this shortcut for Meek elections is the following: partition the set of candidates into `canonical winners,' `strong,' and `weak' candidates. Then, statistically verify that each of the canonical winners beats each of the weak candidates, as long as the hopeful set contains all of the canonical winners. This can look a lot like the so-called ``Never Loses (NL)'' assertions in \textcite{blom_3_2025}; or similarly we could compare the winner's first place votes to the weak candidate's mentions. If we can show this, then we are certain that all of the weak candidates must be eliminated before all of the canonical winners, and we know that the only possible vertices after the elimination cloud include all of the canonical winners and some subset of the strong candidates; by considering every such subset, we can initialize the graph in a `post-elimination cloud' state.\4 
One mathematical simplification of our framework would be to find a way to deal with the parameter explosion we expect for higher degree local nulls by simultaneously bounding the effect of all high-degree transfers. As seen in Equation (\ref{eq:degree_2}), there are already 17 parameters involved in the constuction of degree-2 local null audits. Even if Proposition \ref{prop:ASN} shows that the ASNs of our audits are only proportional to the square root of this number $d$ of parameters, we expect the number of parameters to blow up combinatorially with degree. To keep ASNs reasonable, it will therefore be necessary either to 1) improve the hypergeometric variance bounding strategy outlined in Section \ref{sec:variance_bounds} (which is very plausible), or 2) find some way of combining the effect of all higher-degree transfers into a single worst-case scenario parameter, to avoid tracking all the high-degree effects individually. A priori, there is reason to believe this to be something we can accomplish -- intuitively, higher-degree transfers have the least impact on margin uncertainty, because votes that transfer through more winners lose more of their weight. But finding the right mathematical language to justify this may be tricky.\4 
Still on the subject of mathematical work, it would be good to have a better description of the properties needed of the profile projections for an election state to be `sufficiently far away' from degeneracy. Our statistical framework for Meek auditing is only justified as long as the instant keep factor solutions to the system of equations $\vec{F}$ in (\ref{eq:vector_form}) live in a connected component of the solution space where the jacobian of $\vec{F}$ is non-singular. Describing (and testing) necessary conditions for this criterion to be fulfilled would make this auditing framework more airtight. The best we can currently do is verify that at least one of the winners in all degree-2 election states we include in our audit graph is secure; in such a case it is impossible for the election state to become degenerate. But a better procedure would surely be needed to audit elections with more seats.\4
Finally, one interesting line of thought might be to consider alternative election rulesets that combine the best of both worlds from Meek and WIGM STV. The advantage of Meek is its structural independence from chronology, whereas the WIGM rule's advantage is computational simplicity. Both of these traits might reasonably be combined into a `hybrid' election rule: for example, by re-computing transfer values/keep-factors each round as Meek does, but only running one iteration of the calibration process, such that there is no need to consider their convergence. Such a modification will surely have to consider chronology in some limited capacity -- to account for scenarios where the second winner only makes quota because of transfers from the first, it will at least be necessary to calibrate the winners' transfer values in the order they got elected, meaning election states will have to track this election order. This would still have an advantage over WIGM in that it would not be necessary to track the set of hopeful candidates at the time of each winner's election. In the Portland District 4 example, this would make it so that all 4,097,797 edges that saw Olivia Clark elected could still collapse down to the same election state in time. \4
The re-calculation of quota in every round, though elegant, might also best be avoided to keep a hybrid rule simple. Such a modified STV rule would still be flexible enough for the audit graphs not to diverge in cases such as our case study, while avoiding the computational challenges we noted in Section \ref{sec:keep} -- the dependence of the transfer values on the profile projections would be cleaner, and it would certainly remove the possibility of degenerate election states.
\printbibliography
\appendix
\section{Audit Graph for Portland D1}\label{sec:portland_d1}
\centering
\includegraphics[width=.6\textwidth]{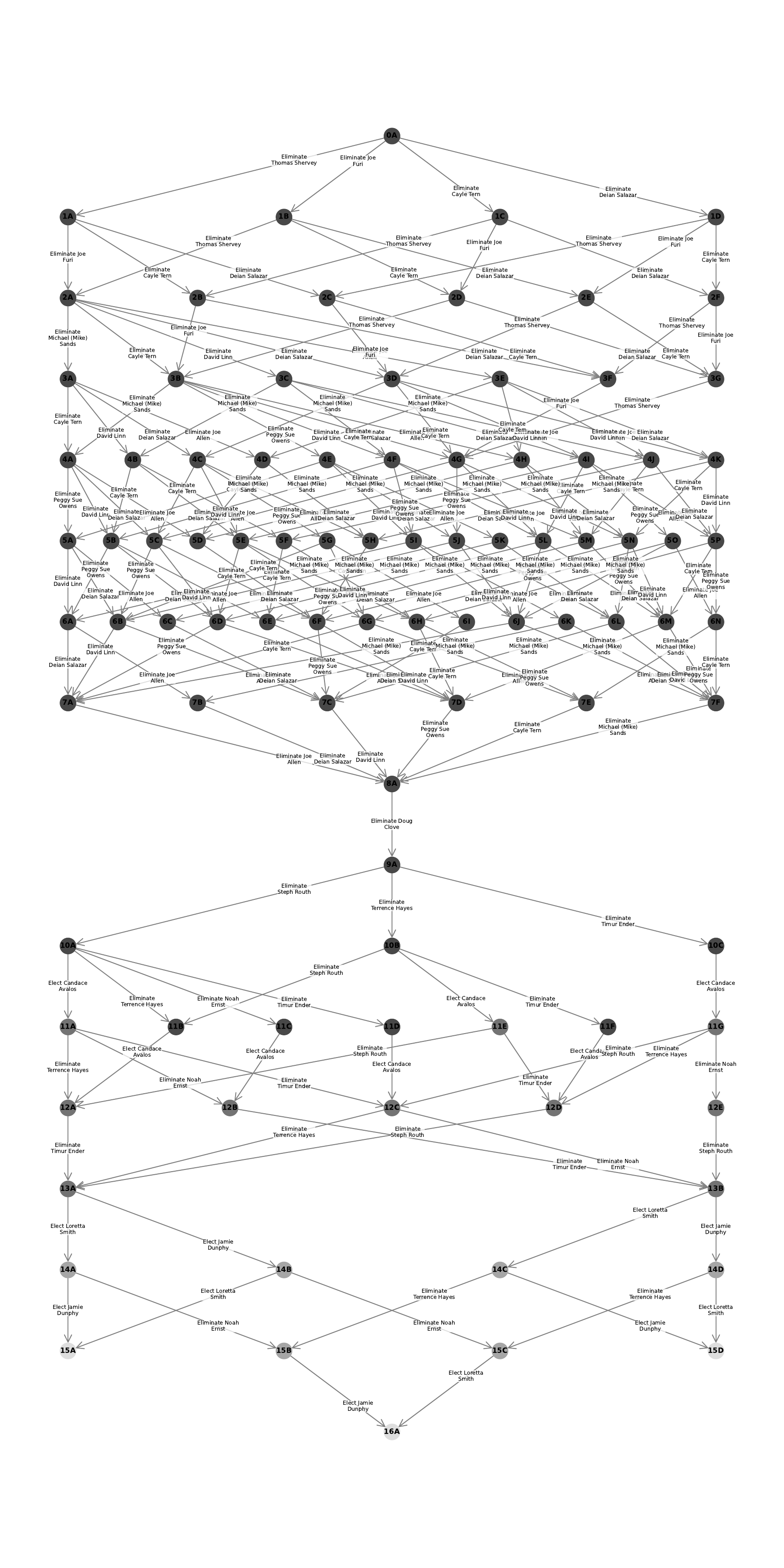}
\end{document}